\documentclass[11pt,a4paper]{article}
\pdfoutput=1
\usepackage{jcappub}
\usepackage{dcolumn}
\usepackage{multirow}

\makeatletter
\DeclareRobustCommand*{\bfseries}{%
  \not@math@alphabet\bfseries\mathbf
  \fontseries\bfdefault\selectfont
  \boldmath
}
\makeatother

\begin{document}

\newcommand{\bb}{\ensuremath{\beta\beta}}
\newcommand{\bbonu}{\ensuremath{\beta\beta0\nu}}
\newcommand{\bbtnu}{\ensuremath{\beta\beta2\nu}}
\newcommand{\Monu}{\ensuremath{\Big|M_{0\nu}\Big|}}
\newcommand{\Mtnu}{\ensuremath{\Big|M_{2\nu}\Big|}}
\newcommand{\Gonu}{\ensuremath{G_{0\nu}(\Qbb, Z)}}
\newcommand{\Gtnu}{\ensuremath{G_{2\nu}(\Qbb, Z)}}

\newcommand{\mbb}{\ensuremath{m_{\beta\beta}}}
\newcommand{\kgy}{\ensuremath{\rm kg \cdot y}}
\newcommand{\ckky}{\ensuremath{\rm counts/(keV \cdot kg \cdot y)}}
\newcommand{\mbba}{\ensuremath{m_{\beta\beta}^a}}
\newcommand{\mbbb}{\ensuremath{m_{\beta\beta}^b}}
\newcommand{\mbbt}{\ensuremath{m_{\beta\beta}^t}}
\newcommand{\nbb}{\ensuremath{N_{\beta\beta^{0\nu}}}}

\newcommand{\Qbb}{\ensuremath{Q_{\beta\beta}}}

\newcommand{\Tonu}{\ensuremath{T_{1/2}^{0\nu}}}

\newcommand{\Ttnu}{\ensuremath{T_{1/2}^{2\nu}}}

\newcommand{\Xe}{\ensuremath{^{136}}Xe}

\newcommand{\Bi}{\ensuremath{^{214}}Bi}

\newcommand{\Tl}{\ensuremath{^{208}}Tl}

\newcommand{\Pb}{\ensuremath{^{208}}Pb}
\newcommand{\PBD}{\ensuremath{^{210}}Pb}

\newcommand{\Po}{\ensuremath{^{214}}Po}

\newcommand{\bru}{cts/(keV$\cdot$kg$\cdot$y)}

\newcommand{\minitab}[2][l]{\begin{tabular}{#1}#2\end{tabular}}

\newcommand{\thedraft}{0.1.1}

\newcommand{\MO}{\ensuremath{{}^{100}{\rm Mo}}}
\newcommand{\SE}{\ensuremath{{}^{82}{\rm Se}}}
\newcommand{\ZR}{\ensuremath{{}^{96}{\rm Zr}}}
\newcommand{\KR}{\ensuremath{{}^{82}{\rm Kr}}}
\newcommand{\ND}{\ensuremath{{}^{150}{\rm Nd}}}
\newcommand{\XE}{\ensuremath{{}^{136}\rm Xe}}
\newcommand{\GE}{\ensuremath{{}^{76}\rm Ge}}
\newcommand{\GES}{\ensuremath{{}^{68}\rm Ge}}
\newcommand{\TE}{\ensuremath{{}^{128}\rm Te}}
\newcommand{\TEX}{\ensuremath{{}^{130}\rm Te}}
\newcommand{\TL}{\ensuremath{{}^{208}\rm{Tl}}}
\newcommand{\CA}{\ensuremath{{}^{48}\rm Ca}}
\newcommand{\CO}{\ensuremath{{}^{60}\rm Co}}
\newcommand{\PO}{\ensuremath{{}^{214\rm Po}}}
\newcommand{\U}{\ensuremath{{}^{235}\rm U}}
\newcommand{\CT}{\ensuremath{{}^{10}\rm C}}
\newcommand{\BE}{\ensuremath{{}^{11}\rm Be}}
\newcommand{\BO}{\ensuremath{{}^{8}\rm Be}}
\newcommand{\UDTO}{\ensuremath{{}^{238}\rm U}}
\newcommand{\CD}{\ensuremath{^{116}{\rm Cd}}}
\newcommand{\THO}{\ensuremath{{}^{232}{\rm Th}}}
\newcommand{\BI}{\ensuremath{{}^{214}}Bi}
\newcommand{\RN}{\ensuremath{{}^{222}}Rn}

\title{Sense and sensitivity of double beta decay experiments}

\author[a]{J.J.~G\'omez-Cadenas,} \emailAdd{gomez@mail.cern.ch}
\author[a]{J.~Mart\'in-Albo,} \emailAdd{justo.martin-albo@ific.uv.es}
\author[a]{M.~Sorel,} \emailAdd{sorel@ific.uv.es}
\author[a]{P.~Ferrario,} \emailAdd{paola.ferrario@ific.uv.es}
\author[a]{F.~Monrabal,} \emailAdd{francesc.monrabal@ific.uv.es}
\author[a]{J.~Mu\~noz,} \emailAdd{jmunoz@ific.uv.es}
\author[b]{P.~Novella,} \emailAdd{pau.novella@ciemat.es}
\author[c]{A.~Poves} \emailAdd{alfredo.poves@uam.es}

\affiliation[a]{Instituto de F\'isica Corpuscular (IFIC), CSIC \& Universidad de Valencia \\
46980 Valencia, Spain}
\affiliation[b]{Centro de Investigaciones Energ\'eticas, Medioambientales y Tecnol\'ogicas (CIEMAT)\\
28040 Madrid, Spain}
\affiliation[c]{Departamento de F\'isica Te\'orica and IFT, Universidad Aut\'onoma de Madrid\\
28049 Madrid, Spain}

\abstract{The search for neutrinoless double beta decay is a very active field in which the number of proposals for next-generation experiments has proliferated. In this paper we attempt to address both the sense and the sensitivity of such proposals. Sensitivity comes first, by means of proposing a simple and unambiguous statistical recipe to derive the sensitivity to a putative Majorana neutrino mass, $m_{\beta\beta}$. In order to make sense of how the different experimental approaches compare, we apply this recipe to a selection of proposals, comparing the resulting sensitivities. We also propose a ``physics-motivated range'' (PMR) of the nuclear matrix elements as a unifying criterium between the different nuclear models. The expected performance of the proposals is parametrized in terms of only four numbers: energy resolution, background rate (per unit time, isotope mass and energy), detection efficiency, and $\beta\beta$ isotope mass. For each proposal, both a reference and an optimistic scenario for the experimental performance are studied. In the reference scenario we find that all the proposals will be able to partially explore the degenerate spectrum, without fully covering it, although four of them (KamLAND-Zen, CUORE, NEXT and EXO) will approach the 50 meV boundary. In the optimistic scenario, we find that CUORE and the xenon-based proposals (KamLAND-Zen, EXO and NEXT) will explore a significant fraction of the inverse hierarchy, with NEXT covering it almost fully. For the long term future, we argue that  $^{136}$Xe-based experiments may provide the best case for a 1-ton scale experiment, given the potentially very low backgrounds achievable and the expected scalability to large isotope masses.}


\arxivnumber{1010.5112}

\maketitle
\flushbottom

\section{Introduction} \label{sec:Introduction} \label{sec:intro}
Physicists invent, build and run experiments to search for new phenomena. In order to convince the funding agencies to support their research, they write proposals in which they estimate the {\em sensitivity} of their experiments.  The definition of sensitivity is somewhat perverse: rather than promising a discovery, the proponents of a new experiment assume that {\em it will fail to find a signal}, and attempt to demonstrate that, in this worst-case scenario, they will  exclude a larger portion of the landscape of physical parameters (such as a cross section or a lifetime) than other proposals. Loosely speaking, the sensitivity allows to compare the physics case of competing experimental approaches. 

An area in which the proposals for new experiments have proliferated during the last decade is the search for neutrinoless double beta decay (\bbonu) --- see, for instance, \cite{Avignone:2007fu, Giuliani:2010} and references therein. The detection of such a process would establish that neutrinos are Majorana particles \cite{Schechter:1981bd, Hirsch:2006yk} (that is to say, truly neutral particles indistinguishable from their antiparticles), and a measurement of the decay rate would provide direct information on neutrino masses \cite{Elliott:2002xe}. The observation of neutrino oscillations \cite{GonzalezGarcia:2007ib} --- which implies that neutrinos have a non-zero mass, an essential condition for \bbonu\ to exist --- and the possible evidence of a \bbonu\ signal in the Heidelberg-Moscow experiment \cite{KlapdorKleingrothaus:2001ke} have boosted the interest in the double beta decay searches, and prompted a new generation of experiments with improved sensitivity. In spite of the formidable experimental challenge (or because of it) the field is rich in new ideas. Among the newcomers one finds many experimental approaches, as well as several source isotopes. All of them claim to be capable of reaching a sensitivity to an {\em effective Majorana neutrino mass} (defined in Sec.~\ref{sec:bb}) of at least 100 meV, and many advertise a second phase in which the sensitivity can be improved to some 20 meV.  

However, considerable confusion can happen because the different proposals often use different recipes to compute the sensitivity. Also, similar sensitivity results hide sometimes more or less credible detector performance assumptions, concerning, for example, energy resolution or radiopurity. An additional complication arises when comparing the sensitivity of experimental approaches using different isotopes. In such a case, a large theoretical uncertainty exists in relating the decay rate sensitivity of a proposal to the effective Majorana neutrino mass sensitivity (see Sec.~\ref{sec:bb}). And to further blur the picture, the scalability, cost, and schedules of the proposed experiments vary greatly. 

In this paper we attempt to address both the sense and the sensitivity of the neutrinoless double beta decay experimental approaches currently proposed. In Section \ref{sec:bb} we discuss the relationship between the rate of \bbonu, neutrino masses and nuclear theory, motivating our assumptions for the latter. Section \ref{sec:exp} briefly reviews the experimental aspects of \bbonu\ searches. In Section \ref{sec:stats} we discuss in detail our definition of sensitivity and show the problems that arise when the observed number of events is small compared to the expected background rate. Next, the Feldman-Cousins unified approach \cite{Feldman:1997qc} is presented and is applied to a \bbonu\ experiment with background (we give more details of this procedure in Appendix \ref{sec:fc}). Section \ref{sec:proposals} describes the main features of the experiments taken into consideration, defining a plausible range of detector performance indicators for each proposal. We try to make sense of these proposals in Section \ref{sec:results}, where the corresponding \mbb\ sensitivities are compared as a function of exposure (kg $\cdot$ year) and for different scenarios. We briefly address also the question of scalability to large isotope masses. 

\section{Rate of neutrinoless double beta decay} \label{sec:bb}
Neutrinoless double beta decay is a very rare nuclear transition that occurs if neutrinos are massive Majorana particles \cite{Schechter:1981bd,Hirsch:2006yk}. It involves the decay of a nucleus with $Z$ protons into a nucleus with $Z+2$ protons and the same mass number $A$, accompanied by the emission of two electrons: $(Z, A) \rightarrow (Z+2, A) + 2\ e^-$. The sum of the kinetic energies of the two emitted electrons is always the same, and corresponds to the mass difference between mother and daughter nuclei, \Qbb. The decay violates lepton number conservation and is therefore forbidden in the Standard Model. 

The simplest underlying mechanism of \bbonu\ is the virtual exchange of light ($m_{\nu_i}<10$ MeV) Majorana neutrinos, although, in general, any source of lepton number violation (LNV) can induce \bbonu\ and contribute to its amplitude. If we assume that the dominant LNV mechanism at low energies is the light-neutrino exchange, the half-life of \bbonu\ can be written as:
\begin{equation}
(T^{0\nu}_{1/2})^{-1} = \Gonu \ \Monu^2 \ \mbb^2 ,
\label{eq:Tonu}
\end{equation}
where \Gonu\ is an exactly-calculable phase-space factor, $|M^{0\nu}|$ is a nuclear matrix element, and \mbb\ is the effective Majorana mass of the electron neutrino:
\begin{equation}
m_{\beta\beta} = \Big| \sum_{i} U^{2}_{ei} \ m_{i} \Big|,
\label{eq:mbb}
\end{equation}
where $m_{i}$ are the neutrino mass eigenstates and $U_{ei}$ are elements of the neutrino mixing matrix.

Neutrino oscillation experiments constrain how the effective Majorana mass in Eq.~\ref{eq:mbb} changes with the absolute neutrino mass scale, defined as $\min \{m_1,m_3 \}$. Only upper bounds on the absolute neutrino mass, of order 1 eV, currently exist. Also, current neutrino oscillation results cannot differentiate between two possible mass orderings, usually referred to as \textit{normal} and \textit{inverted} orderings. In the normal ordering case, the gap between the two lightest mass eigenstates corresponds to the small mass difference, measured by solar experiments. In this case, the effective Majorana mass can be as low as 2 meV \cite{Giuliani:2010}. If the spectrum is extremely hierarchical, and therefore the absolute neutrino mass can be neglected, $m_{\beta\beta}$ can be as high as 5 meV in the normal ordering case. In the inverted ordering case, the gap between the two lightest states corresponds to the large mass difference, measured by atmospheric experiments. In this case, $m_{\beta\beta}$ can be as low as about 15 meV \cite{Giuliani:2010}. The upper limit on $m_{\beta\beta}$ if the neutrino mass can be neglected is approximately 50 meV in this case. Finally, in the particular case in which the neutrino mass differences are very small compared with its absolute scale, we speak of the \textit{degenerate} spectrum. In this case, larger values for $m_{\beta\beta}$ can be obtained, approximately above 50 meV. 

All nuclear structure effects in \bbonu\ are included in the nuclear matrix element (NME). Its knowledge is essential in order to relate the measured half-life to the neutrino masses, and therefore to compare the sensitivity and results of different experiments, as well as to predict which are the most favorable nuclides for \bbonu\ searches. Unfortunately, NMEs cannot be separately measured, and must be evaluated theoretically.
 
\begin{figure}[t!b!]
\centering
\includegraphics[width=0.75\textwidth]{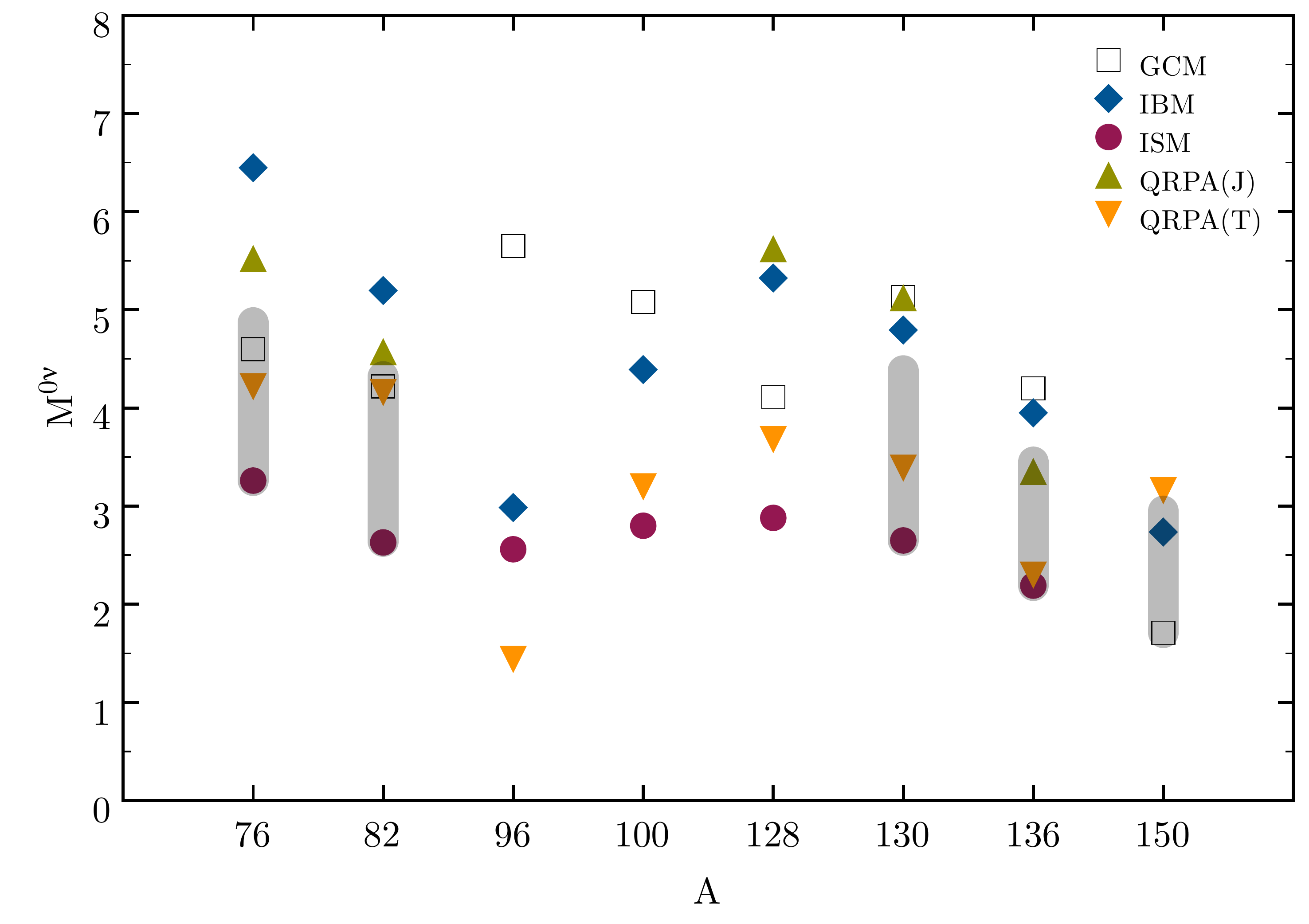}
\caption{Recent NME calculations from the different techniques (GCM \cite{Rodriguez:2010mn}, IBM \cite{Barea:2009zza}, ISM \cite{Menendez:2009di, Menendez:2009oc}, QRPA(J) \cite{Suhonen:2010ci}, QRPA(T) \cite{Simkovic:2009cu,Simkovic:2009fv,Fang:2010qh}) with UCOM short range correlations. All the calculations use $g_{A} = 1.25$; the IBM-2  results are multiplied by 1.18 to account for the difference between Jastrow and UCOM, and the RQRPA are multiplied by 1.1/1.2  so as to line them up with the others in their choice of r$_0$=1.2~fm. The shaded intervals correspond to the proposed physics-motivated ranges (see text for discussion).}
\label{fig.NME}
\end{figure}

In the last few years the reliability of the calculations has greatly improved, with several techniques being used, namely: the Interacting Shell Model (ISM)
\cite{Caurier:2007wq, Menendez:2009di, Menendez:2009oc}; the Quasiparticle Random Phase Approximation (QRPA) \cite{Suhonen:2010ci, Simkovic:2008fr, Simkovic:2009cu}; the Interacting Boson Model (IBM) \cite{Barea:2009zza}; and the Generating Coordinate Method (GCM) \cite{Rodriguez:2010mn}. Figure \ref{fig.NME} shows the most recent results of the different methods. We can see that in most cases the results of the ISM calculations are the smallest ones, while the largest ones may come from the IBM, QRPA or GCM. 

Shall the differences between the different methods be treated as an uncertainty in sensitivity calculations? Should we assign an error bar to the distance between the maximum and the minimum values? This approach has, we argue, the undesirable effect of blurring the relative merits of different experimental approaches, and does not reflect the recent progress in the theoretical understanding of the treatment of nuclear matrix elements.

Each one of the major methods has some advantages and drawbacks, whose effect in the values of the NME can be sometimes explored. The clear advantage of the ISM calculations is their full treatment of the nuclear correlations, while their drawback is that they may underestimate the NMEs due to the limited number of orbits in the affordable valence spaces. It has been estimated \cite{Blennow:2010th} that the effect can be of the order of 25\%. On the contrary,  the QRPA variants, the GCM in its present form, and the IBM are bound to  underestimate the multipole correlations in one or another way.  As it is well established that these  correlations tend to diminish the NMEs, these methods should tend to overestimate them \cite{Caurier:2007wq, Menendez:2010id}.

With these considerations in mind, we propose here our best estimate to NME values obtained as the central value of a physics-motivated range (PMR) of theoretical values for \GE, \SE, \TEX, \XE\ and \ND. In what follows we select the results of the major nuclear structure approaches which share the following common ingredients: (a) nucleon form factors of dipole shape; (b) soft short range correlations computed with the UCOM method \cite{Feldmeier:1997zh}; (c) unquenched axial coupling constant g$_{\rm A}$; (d) higher order corrections to the nuclear current \cite{Simkovic:1999re}; and (e) nuclear radius $R=r_0\ A^{1/3}$, with $r_0=1.2$~fm \cite{Smolnikov:2010zz}. Therefore, the remaining discrepancies between the diverse approaches are solely due to the different nuclear wave functions that they employ. 

Let's start with the \ND\ case, for which no ISM value is available. The GCM calculation \cite{Rodriguez:2010mn} is clearly the most sophisticated in the market from the point of view of the nuclear structure, and gives the smaller NME. The two other approaches, QRPA \cite{Fang:2010qh} and IBM \cite{Barea:2009zza}, give larger and similar results; therefore we average them both to propose a NME range $[1.71-2.95]$, and therefore a NME best estimate of $2.33$.

For \XE\ we have an ISM value which defines the lower end of the range \footnote{We could have increased this value in view of the above discussion, but we shall refrain for doing so and use just published values.}. For the upper one, we average the NMEs from the RQRPA calculation of the T\"ubingen group \cite{Simkovic:2009cu}, the GCM \cite{Rodriguez:2010mn}, the IBM  and the more recent pnQRPA result from the Jyv\"askyl\"a La Plata collaboration \cite{Suhonen:2010ci}. The resulting interval is $[2.19-3.45]$, and our most probable value is $2.82$.  

With the same ingredients we obtain a range $[2.65-4.61]$ for \TEX , corresponding to a best estimate of $3.63$. 

For \SE\ the interval is $[2.64-4.32]$ using the latest SRQRPA results \cite{Simkovic:2009fv}, and the NME most probable value is $3.48$. 
 
Finally we come to \GE\ where we can use an extra filter, namely, to demand that the calculations must be consistent with the occupation numbers measured by J.~Schiffer and collaborators \cite{Schiffer:2007ut, Kay:2008dg}. This leaves us with the ISM \cite{Menendez:2009oc}, the SRQRPA, \cite{Simkovic:2009fv} and the pnQRPA \cite{Suhonen:2010ci}. Averaging again the two QRPA values we obtain the interval $[3.26-4.87]$, and a best estimate of $4.07$. 

The properties of the five isotopes relevant to our work, including our proposed values for the NMEs, are shown in Table \ref{tab:nme}. For the sake of simplicity and for the above-mentioned intended focus on directly comparing different experimental approaches, the majority of the results in this work rely only on our best estimate for NME values. On the other hand, we acknowledge that some uncertainty in NME calculations still exist, and that our recipe for estimating NME most probable values is by no means unique. For this reason, we also tabulate in Sec.~\ref{subsec:scalability} our final \mbb\ results separately for all NME calculations considered above.

\begin{table}[t!b!]
\centering
{\small
\begin{tabular}{rD{.}{.}{-1}ccccc}
\hline\hline \noalign{\smallskip}
Isotope & \multicolumn{1}{c}{$W$}  &  \multicolumn{1}{c}{\Qbb}  &  \multicolumn{1}{c}{$|M_{0\nu}|$}  &  \multicolumn{1}{c}{$|G_{0\nu}|^{-1}$}  &  \multicolumn{1}{c}{$T^{0\nu}_{1/2}(\mbb=50\ \mathrm{meV})$} & \multicolumn{1}{c}{$N_{0\nu}/N_{0\nu}(\mathrm{Ge})$}   \\[1pt]
& \multicolumn{1}{c}{$(\mathrm{g/mol})$}   &  \multicolumn{1}{c}{(keV)}  &  \multicolumn{1}{c}{}  &  \multicolumn{1}{c}{($10^{25}$ y~eV$^{2}$)} & \multicolumn{1}{c}{$(10^{27}\ \mathrm{y})$} & \\ \hline \noalign{\smallskip}
\GE  &  75.9 & 2039 & 4.07 & 4.09  & 0.95 & 1.0 \\
\SE  &  81.9 & 2996 & 3.48 & 0.93  & 0.26 & 3.3 \\
\TEX & 129.9 & 2528 & 3.63 & 0.59  & 0.18 & 3.1 \\
\XE  & 135.9 & 2458 & 2.82 & 0.55  & 0.25 & 2.1 \\
\ND  & 149.9 & 3368 & 2.33 & 0.13  & 0.15 & 3.3 \\ 
\hline\hline
\end{tabular}}
\caption{Physical properties of different isotopes considered in this paper: atomic weight, $W$ \cite{Audi:2003}; \bb\ decay $Q$-value \cite{Mount:2010zz, Nxumalo:1993jn, Redshaw:2009zz, Redshaw:2007un, Kolhinen:2010zz}; the PMR for the nuclear matrix element ($|M_{0\nu}|$, see text for discussion); inverse of \bb\ decay phase-space factor ($|G_{0\nu}|^{-1}$, from \cite{Zuber:2005fu}). For illustrative purposes, we also give the \bbonu\ half-life for a fixed \mbb\ value ($T^{0\nu}_{1/2}(\mbb=50\ \mathrm{meV})$), and the expected number of isotope \bbonu\ decays relative to \GE\ \bbonu\ decays for a fixed isotope mass ($N_{0\nu}/N_{0\nu}(\mathrm{Ge})$).}
\label{tab:nme}
\end{table}

\section{Experimental aspects} \label{sec:exp}
The choice among \bb\ isotopes discussed in Sec.~\ref{sec:bb} is not the only important factor to be considered when optimizing future \bbonu\ proposals. According to Table \ref{tab:nme}, one should prefer, everything else being the same, experiments based on \ND, \SE, \TEX, or even \XE, rather than experiments based on \GE. However, Ge-based experiments have dominated the field so far.

The reason is that \bbonu\ experiments must be designed to measure the kinetic energy of the electrons emitted in the decay. Due to the finite energy resolution of any detector, \bbonu\ events are reconstructed within a non-zero energy range centered around \Qbb, typically following a gaussian distribution. As will be demonstrated in Sec.~\ref{sec:results}, any background event falling in this energy range limits dramatically the sensitivity of the experiment. Good energy resolution is therefore essential. Germanium semiconductor detectors provide the best energy resolution achieved to date: in a \GE\ experiment, a region able to contain most of the signal ---  called the \emph{region of interest} (ROI), and often taken as 1 FWHM around \Qbb\ --- would be only a few keV wide.  

Unfortunately, energy resolution is not enough by itself: a continuous spectrum arising from natural decay chains can easily overwhelm the signal peak, given the enormously long decay times explored. Consequently, additional signatures to discriminate between signal and backgrounds, such as event topology or daughter ion tagging, are desirable. Also, the experiments require underground operation and a shielding to reduce external background due to cosmic rays and surrounding radioactivity, and the use of very radiopure materials. In addition, large detector masses, high \bb\ isotope enrichment, and high \bb\ detection efficiency are clearly desirable, given the rare nature of the process searched for. No experimental technique scores the highest mark in all of the above, and thus different approaches are possible. 

The Heidelberg-Moscow (HM) experiment \cite{Klapdor:2000sn}, using high-purity germanium diodes enriched to 86\% in the isotope \GE, set the most sensitive limit to date: $T^{0\nu}_{1/2}(\GE) \ge 1.9 \times 10^{25}$ years (90\% CL). The experiment accumulated a total exposure of 71.7 \kgy, and achieved a background rate in the ROI of 0.06 \ckky\ after pulse shape identification. The energy resolution (FWHM) at \Qbb\ was $4.23 \pm 0.14$ keV. A subset of the Collaboration observed evidence for a \bbonu\ signal \cite{KlapdorKleingrothaus:2001ke}. The claim has been severely questioned \cite{Aalseth:2002dt}, but no one has been able to prove it wrong. According to it, the isotope \GE\  would experiment \bbonu\ decay with a lifetime of about $1.5\times10^{25}$ years. Using the PMR nuclear matrix element, this corresponds to a neutrino mass of about 0.4 eV.

In this paper, we compare the \mbb\ sensitivity of different experimental approaches by parametrizing in a simple and intuitive way their detector performance, as will be discussed in Sec.~\ref{sec:proposals}. Before that, in Sec.~\ref{sec:stats}, we discuss the statistical recipe that we will adopt to derive the experimental sensitivities.  

\section{A unified treatment of sensitivity estimates of \bbonu\ experiments} \label{sec:stats}
\subsection{An ideal experiment} \label{sec:ideal}
All \bbonu\ experiments have to deal with non-negligible backgrounds, an only partially efficient \bbonu\ event selection, and more or less difficulties to extrapolate their detection technique to large masses. It is instructive, however, to imagine an ideal experiment defined by the following parameters: (a) the detector mass is 100\% made of a \bb\ source; (b) perfect energy resolution, and/or infinite radiopurity, resulting in a null background rate; (c) perfect detection efficiency; and (d) scalable to large masses at will, that is, exposure as large as desired. 

Suppose that such an experiment runs for a total exposure $Mt$. The expected number of \bbonu\ events is given by
\begin{equation}
N = \log2 \cdot \frac{N_{A}}{W} \cdot \varepsilon \cdot \frac{M\cdot t}{T_{1/2}^{0\nu}}, \label{eq.N}
\end{equation}
where $N_{A}$ is the Avogadro constant, $W$ is the atomic weight of the \bb-decaying isotope, $\varepsilon$ is the signal detection efficiency, $M$ is the \bb\ isotope mass and $t$ the data-taking time.

\begin{figure}[b!]
\centering
\includegraphics[width=0.475\textwidth]{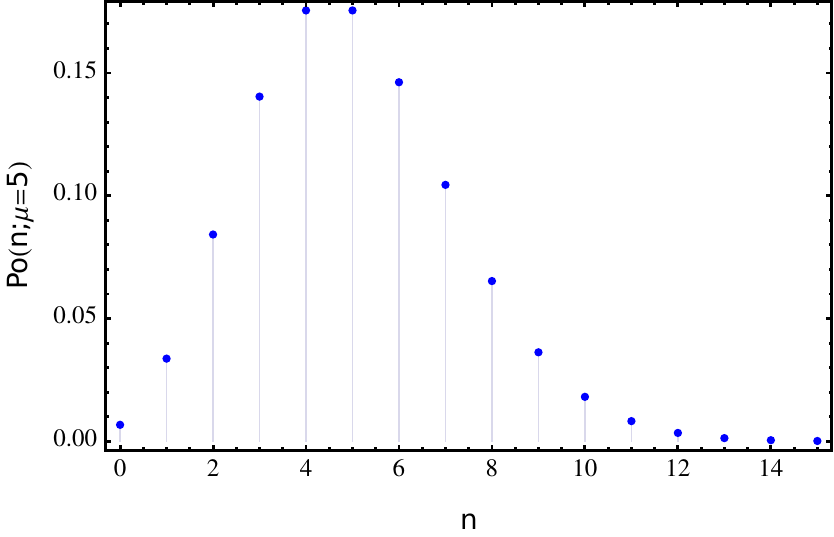}
\includegraphics[width=0.475\textwidth]{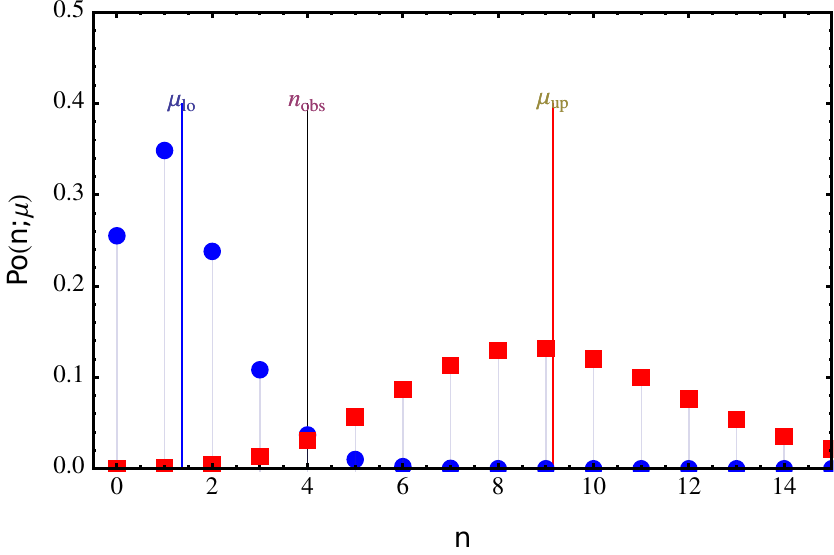}
\caption{Left: a Poisson distribution with $\mu=5$. The probability of observing 4 events is 17.5\%, which means that only that fraction of experiments will yield a number of events equal to the expectation value $\mu$. Right: 
lower ($\mu_{\rm lo}$) and upper ($\mu_{\rm up}$) limits  computed for an observed number of events $n_{\rm obs}$ equal to 4, and $\alpha=\beta=0.05$. Two Poisson distributions are shown: ${\rm Po}(n;\mu =\mu_{\rm lo})$ and ${\rm Po}(n;\mu =\mu_{\rm up})$.}
\label{fig.PoisonMu5}
\end{figure}

Different identical experiments running for the same total exposure would observe a different number of events, $n$, with expectation value $\mu = N$. The probability distribution function (pdf) of $n$ is a Poisson distribution: 
\begin{equation}
{\rm Po}(n;\mu)=\frac{\mu^n}{n!} \ e^{-\mu}\,.
\label{eq.po}
\end{equation}
For example, Figure \ref{fig.PoisonMu5} (left) shows a Poisson distribution with $\mu=5$. If we run a large number of identical experiments, only 17.5\% will observe 5 events. In fact, the same percentage will observe 4 events, and a small but non-null number of experiments (7 in a thousand) will observe zero events. Consider now an experiment whose outcome is $n_{\rm obs}=4$. What can be inferred about the true value $\mu$ from this measurement?

A common way to report a result on $\mu$, proposed by J.~Neyman in 1937 \cite{Neyman:1937}, is to give a \emph{confidence interval} (CI) where $\mu$ is likely to be included. How likely an interval is to contain the true value is determined by the \emph{confidence level} (CL), usually expressed as a percentage. The end points of the CI, called the \emph{confidence limits}, are defined in terms of a probability: we define the lower (upper) limit $\mu_{\rm lo}$ ($\mu_{\rm up}$) as the value of the parameter $\mu$ such that, if we carry a large number of experiments following the Poisson distribution ${\rm Po}(n; \mu_{\rm lo})$ (${\rm Po}(n; \mu_{\rm up})$), then a fraction $\alpha$ ($\beta$) of them at most will yield a number larger (smaller) than or equal to $n_{\rm obs}$. Therefore, to compute the limits the following equations can be solved numerically:
\begin{eqnarray}
\alpha =& \sum_{n_{\rm obs}}^{\infty} {\rm Po}(n; \mu_{\rm lo}) \label{eq.alpha} \\
\beta  =& \sum_{0}^{n_{\rm obs}} {\rm Po}(n; \mu_{\rm up}) \label{eq.beta}
\end{eqnarray}

Figure \ref{fig.PoisonMu5} (right) illustrates the above definition. The lower and upper limits are computed for $n_{\rm obs}=4$~and $\alpha=\beta=0.05$. The filled (blue) circles represent  ${\rm Po}(n; \mu_{\rm lo}=1.37)$, while the filled (red) squares represent ${\rm Po}(n; \mu_{\rm up}=9.15)$. Adding the filled circles corresponding to $n<4$ yields a probability of 0.95 (therefore, the integrated probability to find \mbox{$n \ge n_{\rm obs} = 0.05$}). Adding the filled squares corresponding to $n>4$ yields a probability of 0.95 (therefore, the integrated probability to find $n \le n_{\rm obs} = 0.05$). Notice  that our choice of a symmetric interval to \emph{cover} the true value $\mu=5$ is not especially well motivated. We could have decided to set asymmetric intervals, for example, $\alpha=0.07$ and $\beta=0.03$~and would still have a 90\% CL. Nevertheless, this is the most common choice for two-sided confidence intervals.

Consider now another experiment: an unlucky team runs under identical conditions than those of the previous experiment, but finds $n_{\rm obs} = 0$. This is an unlikely but by no means impossible outcome, with a probability of 0.7\%. What should the team report? Clearly a lower limit cannot be found, but an upper limit can still be determined by setting $n_{\rm obs} = 0$ in Eq.~(\ref{eq.beta}):
\begin{equation}
\mu_{\rm up} = -\log (\beta)
\label{eq.upf}
\end{equation}
For $\beta=0.1$ (that is, for 90\% CL), Equation \eqref{eq.upf} yields $\mu_{\rm up}\simeq2.3$, and for $\beta=0.05$ (95\% CL), we find $\mu_{\rm up}\simeq3$. The meaning of the upper limit is very clear. In a Poisson distribution with $\mu=2.3$, one observes 10\% of the times $n=0$, while for $\mu=3.0$, one observes 5\% of the times $n=0$.

Therefore, an ideal experiment that observes no events after a given exposure would report a {\em negative} result, that is, an upper limit in the \bbonu\ decay rate $(T^{0\nu}_{1/2})^{-1}$, or possibly in the more relevant physical parameter \mbb. From Equations (\ref{eq:Tonu}) and (\ref{eq.N}), the latter upper limit can be written as:
\begin{equation}
\mbb = K_{1} \sqrt{\frac{N}{\varepsilon Mt}}
\label{eq.mbbx4}
\end{equation}
where $K_{1}$ is a constant that depends only on the isotope type. Substituting $N$ by 2.3 (3), one gets the upper limit at 90\% (95\%) CL, which improves with $\sqrt{1/Mt}$. 

\begin{figure}[t!b!]
\centering
\includegraphics[width=0.5\textwidth]{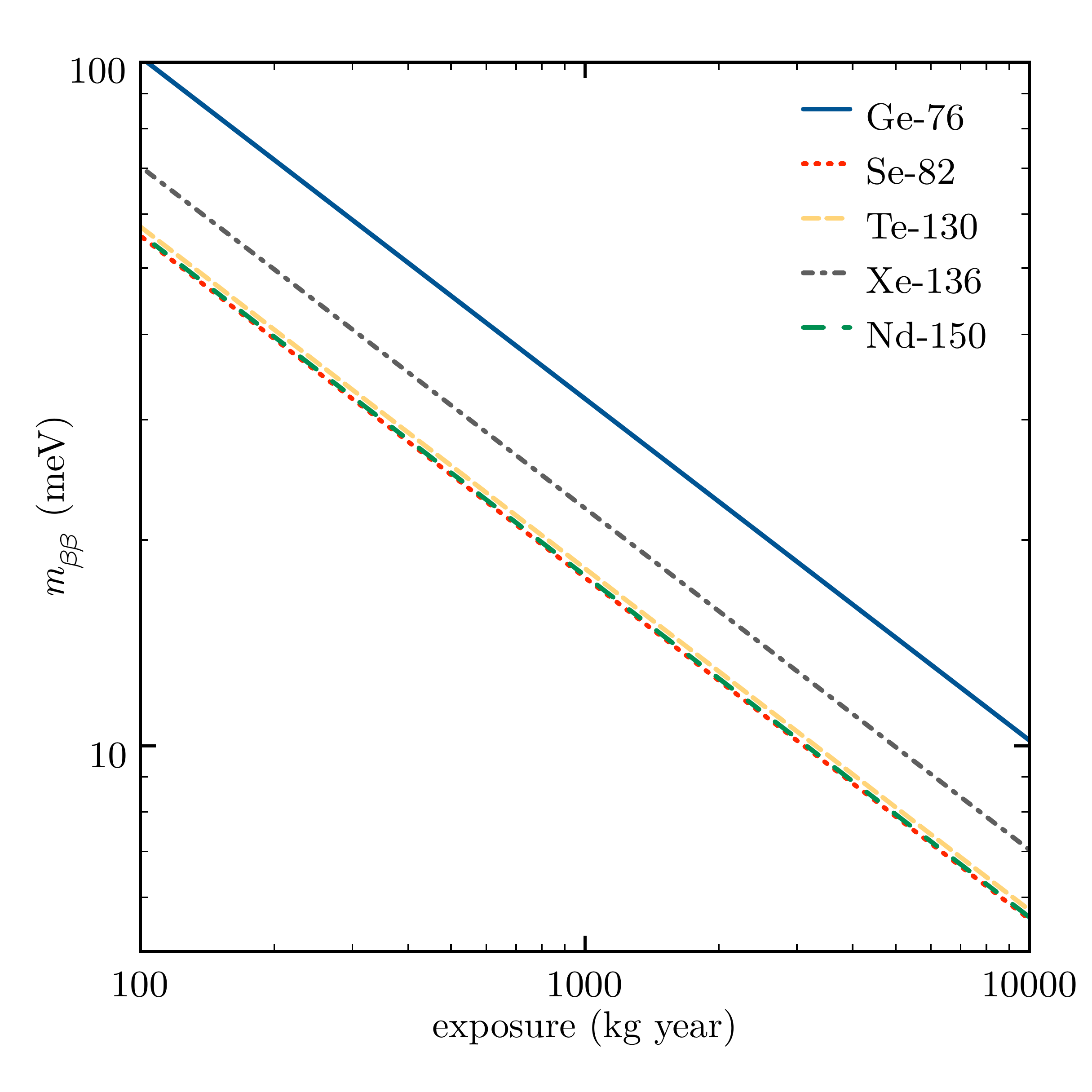}
\caption{
Sensitivity of ideal experiments at 90\% CL for different \bb\ isotopes. Since the yields are very similar, the sensitivities of \SE, \TEX\ and \ND\ overlap.}
\label{fig.snobkg}
\end{figure}

In the following, we define the sensitivity of our experiments as
{\em the average upper limit one would get from an ensemble of experiments with
the expected background and no true signal}.

For an ideal experiment, the expected background is exactly $b=0$, resulting in no Poisson fluctuations in the observed number of signal plus background events for the experiments within the ensemble, which has to be always equal to zero in this case. Therefore the sensitivity of the experiment is in this case simply given by the upper limit reported by the unlucky experiment. Figure \ref{fig.snobkg} shows the sensitivity of the five isotopes considered. Even ideal experiments need a large exposure ($>$10$^3$ \kgy) to fully explore the inverse hierarchy. To start exploring the normal hierarchy, one would need of the order of $10^4$ \kgy, that is, perfect detectors of 1 ton running for 10 years: a truly formidable experimental challenge.  \label{subsec:stats_nobgr}
\subsection{Experiments with background and the collapse of the classical limit}
Consider now the more realistic case of an experiment with background. Call $b$ the expected value of the background, and assume (unrealistically) that it is known with no uncertainty. The relevant pdf will then be: 
\begin{equation}
{\rm Po}(n;\mu+b) = \frac{(\mu+b)^n}{n!} \ e^{-(\mu+b)}\,, 
\label{eq.pobkg}
\end{equation}
where $\mu$ is the (unknown) mean signal expectation and $b$ is the known mean background expectation. The Poisson variable $n$ is such that $n = n_s + n_b$, where the signal and background Poisson variables $n_s$ and $n_b$ have mean expectation values E$[n_s]=\mu$, and E$[n_b]=b$, respectively.

A priori, it appears as if we could treat the problem in exactly the same way as for the case without background. A CI can be constructed using the following equations:
\begin{eqnarray}
\alpha =&  \sum_{n_{obs}}^{\infty} \frac{(\mu_{{\rm lo}} + b)^n}{n!} \ e^{-(\mu_{{\rm lo}} +b)} \label{eq.polimits_lo} \\
\beta =&  \sum_{0}^{n_{obs}} \frac{(\mu_{{\rm up}} + b)^n}{n!} \ e^{-(\mu_{{\rm up}} +b)} \label{eq.polimits_up}
\end{eqnarray} 

Let us work out an explicit example. S{\rm up}pose that we run an experiment in which the predicted background is $b=5$, and we observe $n_{\rm obs}=5$ events. Then, Equation \eqref{eq.polimits_up} becomes: 
\begin{equation}
\beta = \sum_{n=0}^{5} \frac{(\mu_{{\rm up}} + 5)^n}{n!} \ e^{-(\mu_{{\rm up}} +5)}.
\end{equation}
Solving numerically this equation with $\beta=0.05$~(95\% CL) yields an {\rm up}per limit of $\mu_{{\rm up}}=5.51$. The lower limit is also readily computed, obtaining $\mu_{{\rm lo}}=-3.03$. Since $\mu$ is physically bounded to non-negative values, we set $\mu_{{\rm lo}}=0$. 

\begin{figure*}[tb]
\centering
\includegraphics[width=0.45\textwidth]{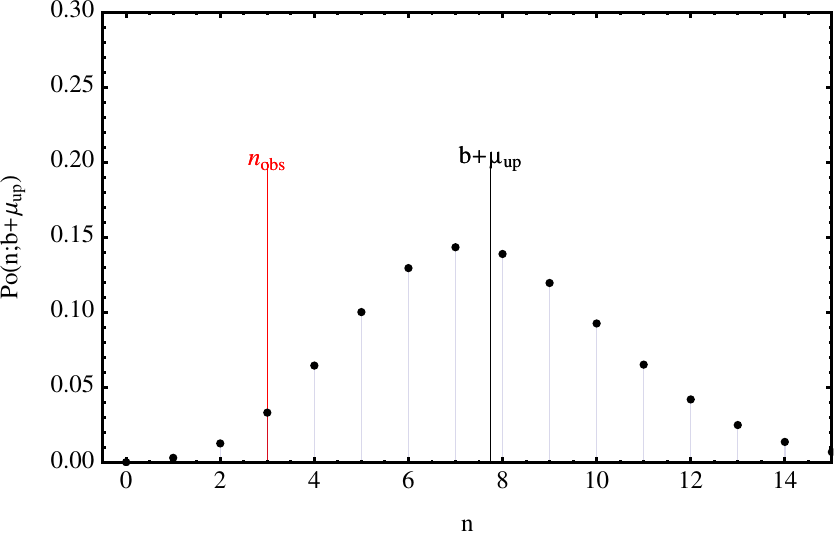}
\includegraphics[width=0.45\textwidth]{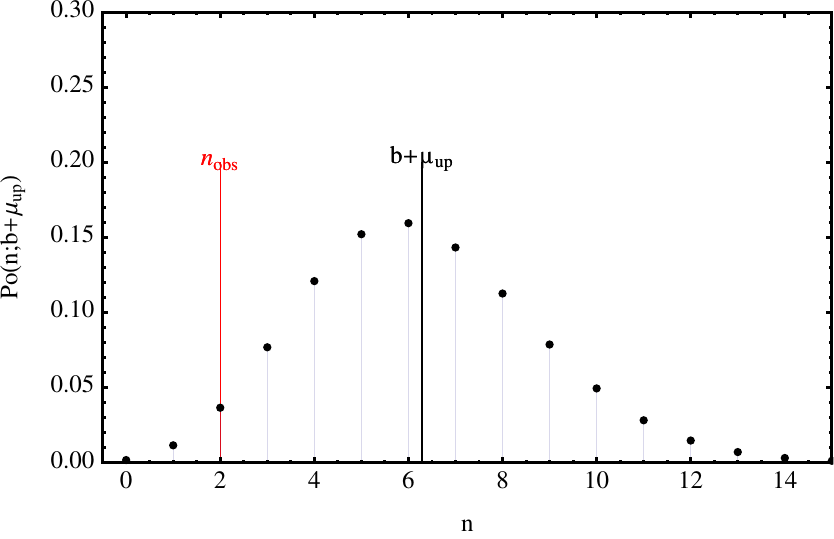}
\includegraphics[width=0.45\textwidth]{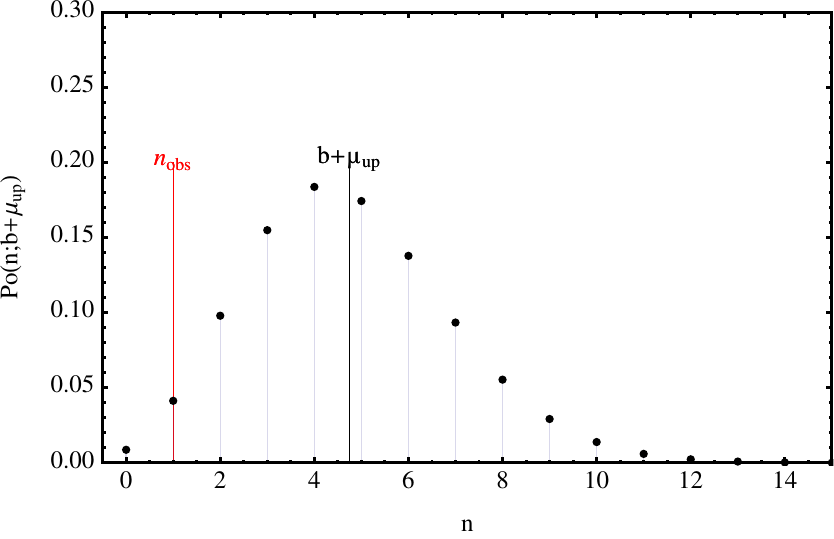}
\includegraphics[width=0.45\textwidth]{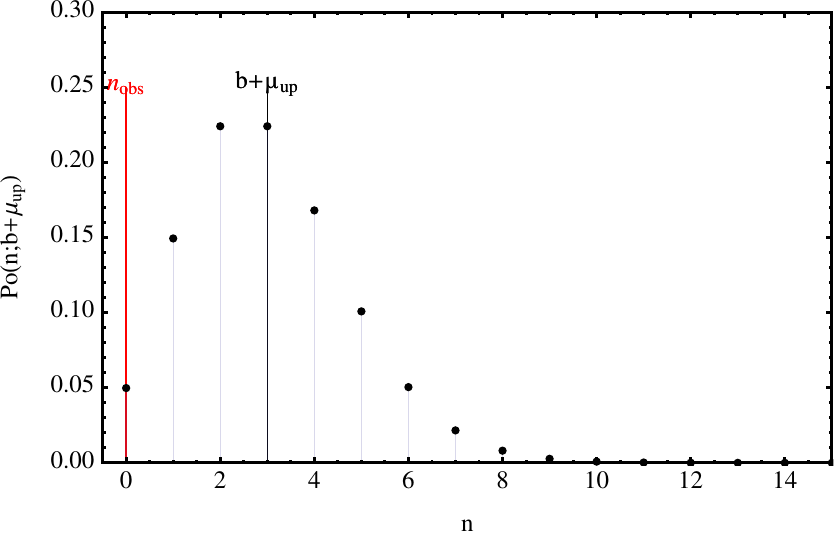}
\caption{The collapse of the classical method: as the number of observed events becomes smaller, the Poisson distribution has to shift left to guarantee that the cumulative percentage to the left of $n_{obs}$ gets to 5\%. In the extreme case of $n_{obs}=0$, the expectation value that yields 5\% of the events in $n=0$, is $\mu + b=3$. Since $b=5$, this forces $\mu_{{\rm up}}=-2$, an absurd result. }
\label{fig.Collapse}
\end{figure*}

\begin{figure}[tb]
\centering
\includegraphics[width=0.6\textwidth]{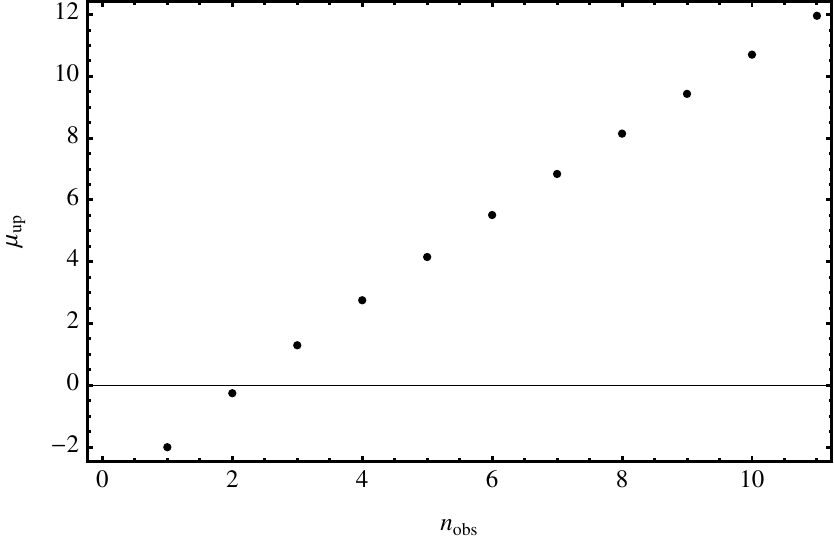}
\caption{Upper limit $\mu_{{\rm up}}$ as a function of the number of observed events $n_{obs}$, estimated solving Eq.~\eqref{eq.polimits_up} for a background prediction $b=5$. The limit does not make sense when the observed number of events is small compared with the expected background ($n_{\rm obs} \ll b$).}
\label{fig.trouble}
\end{figure}

We now repeat the procedure for $n_{obs}=4, 3, 2, 1$ and 0, while keeping $b=5$. In this case, we obtain $\mu_{{\rm up}}=4.15, 2.75, 1.29, -0.25$ and $-2$, respectively. The reason for the decrease in $\mu_{{\rm up}}$ as $n_{obs}$ decreases can be understood by observing Fig.~\ref{fig.Collapse}. As the number of observed events becomes smaller, the Poisson distribution has to shift left to guarantee that the cumulative percentage to the left of $n_{obs}$ gets to 5\%. The situation is summarized in Fig.~\ref{fig.trouble}: due to the discreetness of the Poisson distribution, the ``classical {\rm up}per limit'' method fails to provide meaningful results (i.e., it yields negative {\rm up}per limits) when the observed number of events is small compared with the expected number of background events.  

 \label{subsec:stats_bgr}
\subsection{The unified approach and sensitivity of an experiment with background}
The collapse of the classical limit was solved in 1997 by G.~Feldman and R.~Cousins. In their best-seller paper, ``Unified approach to the classical statistical analysis of small signals'' \cite{Feldman:1997qc}, they introduce a method for the construction of confidence intervals using an ordering principle based on likelihood ratios. The procedure avoids the collapse described in Sec.~\ref{subsec:stats_bgr}, unifying the treatment of upper confidence limits for null results and two-sided confidence limits for non-null results. Since its publication, the unified approach method has become the ``de facto'' standard frequentist approach to compute confidence intervals. We give some details of the method in the Appendix \ref{sec:fc}.

We revisit now the concept of sensitivity in the case of an experiment with background. Recall that by sensitivity we mean the average upper limit that would be obtained by an ensemble of identical replicas of such an experiment, each one with the same mean expected background and no true signal. Let us now translate the above definition into a mathematical formula. We define $\mathcal{U}(n|b)$ as the function yielding the (unified approach) upper limit (at the desired CL) for a given observation $n$ and a mean predicted background level $b$. Values for $\mathcal{U}(n|b)$ are reported in tabular form in \cite{Feldman:1997qc} for several CL values. Given that the variable $n$ follows a Poisson pdf,
${\rm Po}(n|b)\equiv {\rm Po}(n;b)$,
then, according to our definition, the sensitivity $\mathcal{S}(b)$ is given by:
\begin{equation}\
\mathcal{S}(b)\equiv {\rm E}[\mathcal{U}(n|b)]=\sum_{n=0}^{\infty} {\rm Po}(n|b)\ \mathcal{U}(n|b).
\label{eq.Average}
\end{equation}
This equation reads as follows: the sensitivity $\mathcal{S}(b)$ of an experiment expecting $b$ events of background is obtained by averaging the upper limits obtained using the unified approach ($\mathcal{U}(n|b)$) with the likelihood of the individual observations (${\rm Po}(n|b)$). 

\begin{figure}[t!]
\centering
\includegraphics[width=0.475\textwidth]{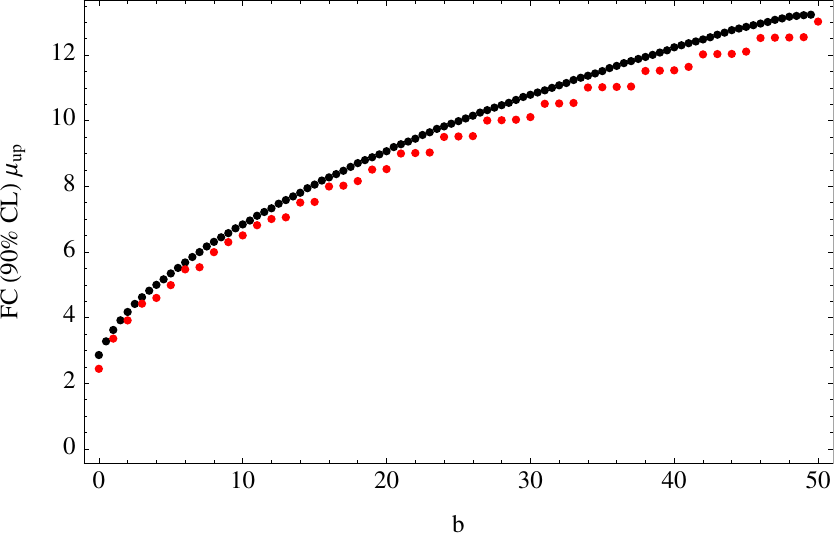}
\includegraphics[width=0.475\textwidth]{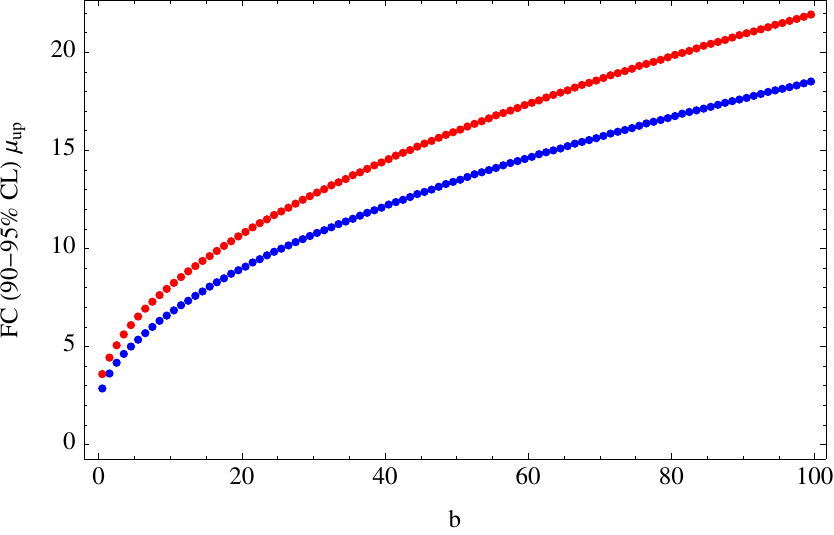}
\caption{Left: 90\% CL sensitivity curve as a function of mean background prediction $b$ as obtained according to Eq.~\eqref{eq.Average} (black dots) or by simply assuming $n=b$ (red dots). Right: sensitivity curve as a function of $b$ for 90\% and 95\% CL.}
\label{fig.FCSensi}
\end{figure}

Let us work out an example. Assume that an experimental team proposes a 100 kg germanium detector to run underground for 10 years. The team claims that the expected background rate in the experiment is $c= 0.001\ \ckky$. Their energy ROI has a width of 5 keV. The predicted background is then $b= 0.001 \times 100 \times 5 \times 10 = 5$~events. If we set $\mu=0$ (no signal) and $b=5$ in Eq.~\eqref{eq.pobkg}, we obtain the same distribution as for the case ($\mu=5$, $b=0$) described in Sec.~\ref{sec:ideal}. If we throw many experiments following ${\rm Po}(n;\mu=5)$, the fraction of the times that we will obtain $n_{\rm obs}=0,1,2,3\ldots$, is therefore described by the pdf shown in the top panel of Fig.~\ref{fig.PoisonMu5}. Consider for instance the case $n=5$. The upper limit at 90\% CL following the prescription of the unified approach is $\mathcal{U}(5|5)= 4.99$ (see \cite{Feldman:1997qc}); we also know that ${\rm Po}(5|5)=0.175$ (see Fig.~\ref{fig.PoisonMu5}). Analogously, for $n=4$, $\mathcal{U}(4|5)= 3.66$ and ${\rm Po}(4|5)=0.175$; for $n=3$, $\mathcal{U}(3|5)= 2.78$ and ${\rm Po}(3|5)=0.14$; and so forth. The sensitivity of the experiment, computed as the average of all the experiments, would then be:
\begin{eqnarray*}
\mathcal{S}(5) & = & \sum_{n=0}^{\infty} {\rm Po}(n|5)\ \mathcal{U}(n|5)  \\*
               & = & 4.99\times0.175 + 3.66 \times 0.175 + \ldots = 5.17
\end{eqnarray*}
This value, $\mathcal{S}(5)=5.17$ (90\% CL) is different than the one sometimes used as upper limit, $\mathcal{U}(5|5)=4.99$. As illustrated in the left panel of Fig.~\ref{fig.FCSensi}, the difference between the $\mathcal{S}(b)$ and $\mathcal{U}(b|b)$ sensitivity estimates is small but significant for all $b$. The latter approach is not strictly correct, introduces fluctuations, and underestimates the sensitivity limit for all $b$. The right panel in Fig.~\ref{fig.FCSensi} compares the 90\% with the 95\% CL sensitivity curves obtained with our sensitivity procedure. 

We should note that, in the large background approximation, the sensitivity curve as a function of $b$ follows the expected classical limit:
\begin{equation}
\mu_{\rm up} \equiv \mathcal{S}(b) \simeq \alpha \cdot \sqrt{b}, \, \, {\rm for\ large\ b},
\label{eq.aupper}
\end{equation}
where $\alpha=1.64$ (1.96) at 90\% (95\%) CL.

If we substitute $N$ in Eq.~\eqref{eq.mbbx4} by the upper limit obtained using the unified approach, then, in the limit of large background:
\begin{equation}
\mbb =K_{2} \sqrt{\frac{b^{1/2}}{\varepsilon Mt}} 
\label{eq.mbbx1}
\end{equation}
where $K_{2}$ is a constant depending on the isotope. If the background $b$ is proportional to the exposure $Mt$ and to the width $\Delta E$ of the ROI:
\begin{equation}
b = c\cdot Mt\cdot \Delta E
\label{eq.mbbx2}
\end{equation}
with the background rate $c$, in \ckky, constant across the energy ROI, then:
\begin{equation}
\mbb = K_2  \ \sqrt{1/\varepsilon} \ \Big(\frac{c \ \Delta E}{Mt}\Big)^{1/4}
\label{eq.mbbx3}
\end{equation}
That is, in the limit of large background, the sensitivity to \mbb\  improves very slowly with exposure, as $(Mt)^{-\frac{1}{4}}$.

In this paper, we compare the sensitivity of several experiments using the unified approach to compute upper limits after Equation \eqref{eq.Average}. As it was the case for our choice of the NME, this is a conservative but robust option. It is conservative because we do not use all the information available to compute upper limits (such as the potentially different energy distributions for the signal and the backgrounds in the ROI). It is robust, however, since the unified approach ensures coverage, unlike, for example, the profile likelihood method in the low background limit. Furthermore, it does not require the precise knowledge of the pdfs of signal and backgrounds in the ROI for the different experiments. While other methods can be used and many have been proposed in the literature (see, for example, \cite{James:2000et}), we believe that the use of the unified approach allows a simple (and therefore easily reproducible) calculation of sensitivities. As discussed in Sec.~\ref{sec:exp} and motivated here in the large exposure limit (but valid for all exposures), proponents of an experiment simply need to provide three parameters describing the detector performance and the isotope to be used, in order to allow the derivation of the \mbb\ sensitivity for a given exposure. The three parameters are: (a) the FWHM energy resolution $\Delta E$, defining the energy ROI (of, say, 1 FWHM around \Qbb ) for the \bbonu\ search; (b) the background rate $c$ per unit of \bb\ isotope mass, energy and time; (c) the \bb\ detection efficiency $\varepsilon$. With this information, the sensitivity can be computed using Equation \eqref{eq.Average}. 
 \label{subsec:stats_fc}

\section{A selection of proposals} \label{sec:proposals}
Many double beta decay experiments have been proposed in the last decade. A recent review \cite{Giuliani:2010} quotes 14 projects in different stages of development and using a variety of detection techniques and isotopes. In this paper we compare the relative merits of 7 of these proposals, chosen as representatives of four different experimental approaches. For each of the 7 experiments considered, we report in Table \ref{tab:proposals} the \bb\ isotope that will (likely) be used, together with the three parameters discussed in Sec.~\ref{subsec:stats_fc} that describe the detector performance. It is possible to find all these figures in the literature, although the level of detail in the discussion and the uncertainties associated with them vary greatly from case to case. 

Two scenarios have been considered regarding the background rates expected in the experiments: the \emph{reference} (R) scenario, which implies in most cases an improvement of at least one order of magnitude over the state of the art represented by Heidelberg-Moscow, and the \emph{optimistic} (O) scenario, which projects a further improvement with respect to the R scenario. In order to evaluate the scalability of the proposals to large masses (see Sec.~\ref{subsec:scalability}), we give also a reference and an optimistic assumption for the \bb\ isotope mass. 

In the following, Sec.~\ref{subsec:proposals_description}, a brief description of the proposals is given. The validity of the background rate assumptions, in light of the results achieved so far, is discussed in Sec.~\ref{subsec:results_b}.

\begin{table}[t!]
\begin{center}
\begin{tabular}{l c c c c D{.}{.}{2.2} r}
\hline\hline
\multirow{2}{*}{Experiment} & \multirow{2}{*}{Isotope} & \multicolumn{1}{c}{Resolution} & \multirow{2}{*}{Efficiency} & & \multicolumn{1}{c}{Background rate} & \multicolumn{1}{c}{Mass}\\
&& \multicolumn{1}{c}{(\% at $\Qbb$)} & & & \multicolumn{1}{c}{($10^{-3}$~cts/(keV kg y))} & \multicolumn{1}{c}{(kg)} \\ \hline \noalign{\smallskip}
\multirow{2}{*}{CUORE} & \multirow{2}{*}{\TEX} & \multirow{2}{*}{0.18} & \multirow{2}{*}{0.80} & R & 40 & 200 \\ 
& & & & O & 1 & 400\\ \hline \noalign{\smallskip}
\multirow{2}{*}{EXO} & \multirow{2}{*}{\XE} & \multirow{2}{*}{3.3} & \multirow{2}{*}{0.70} & R & 1 & 160 \\ 
& & & & O & 0.5 & 1000\\ \hline \noalign{\smallskip}
\multirow{2}{*}{GERDA} & \multirow{2}{*}{\GE} & \multirow{2}{*}{0.16} & \multirow{2}{*}{0.80} & R & 10 & 15 \\ 
& & & & O & 1 & 35 \\ \hline \noalign{\smallskip}
\multirow{2}{*}{KamLAND-Zen} & \multirow{2}{*}{\XE} & \multirow{2}{*}{9.5} & \multirow{2}{*}{0.80} & R & 0.5 & 360 \\ 
& & & & O & 0.1 & 1000\\ \hline \noalign{\smallskip}
\multirow{2}{*}{NEXT} & \multirow{2}{*}{\XE} & \multirow{2}{*}{0.7} & \multirow{2}{*}{0.30} & R & 0.2 & 90 \\ 
& & & & O & 0.06 & 1000\\ \hline \noalign{\smallskip}
\multirow{2}{*}{SNO+} & \multirow{2}{*}{\ND} & \multirow{2}{*}{6.5} & \multirow{2}{*}{0.50} & R & 10 & 50 \\ 
& & & & O & 1 & 500 \\ \hline \noalign{\smallskip}
\multirow{2}{*}{SuperNEMO} & \multirow{2}{*}{\SE} & \multirow{2}{*}{4.0} & \multirow{2}{*}{0.30} & R & 0.4 & 7 \\ 
& & & & O & 0.06 & 100 \\
\hline\hline
\end{tabular}
\end{center}
\caption{Proposals considered in the \mbb\ sensitivity comparison. For each proposal, the isotope that will (likely) be used, together with estimates for detector performance parameters --- FWHM energy resolution, detection efficiency and background rate per unit of energy, time and \bb\ isotope mass --- are given. The efficiencies indicated here do not include the efficiency loss due to the 1 FWHM energy cut, common to all proposals. The background estimates and the \bb\ source mass include both a reference (R) and an optimistic (O) scenario.}\label{tab:proposals} 
\end{table} 

\subsection{Brief description of the proposals}\label{subsec:proposals_description}

The first experimental approach we consider is the \textbf{high-resolution calorimetry}. These are detectors characterized by excellent energy resolution. They have also the advantages of simplicity and compactness. The classic germanium diodes and the bolometers fall in this category.

Two experiments, \textbf{GERDA} \cite{Abt:2004yk, Heider:2007cj, Barabanov:2009zz, Zuzel:2010gz} and \textbf{MAJORANA} \cite{Elliott:2008xc, Guiseppe:2008qu, Henning:2009tt}, will search for \bbonu\ in \GE\ using arrays of high-purity germanium detectors. This is a well-established technique that offers outstanding energy resolution (better than 0.2\% FWHM at the $Q$-value) and high efficiency ($\sim0.80$). In its first phase GERDA expects a background rate of the order of $10^{-2}$ \ckky. The Collaboration aims to improve this by an additional order of magnitude, $10^{-3}$ \ckky\ in the ROI, during its second phase. For its part, MAJORANA anticipates a background rate of $10^{-3}$ counts/(ROI$\cdot$kg$\cdot$y), that is, about 4 times less than the second phase of GERDA. For the sake of concreteness, we choose GERDA to represent the current germanium-based proposals, assigning its first phase --- $10^{-2}$ \ckky\ and 15 kg of isotope mass --- to the R scenario and its second phase --- $10^{-3}$ \ckky\ and 35 kg --- to the O scenario\footnote{The GERDA and MAJORANA Collaborations plan to merge in the future for a \GE\ experiment aiming for a ton-scale isotope mass and for a $10^{-4}$ \ckky\ background rate. Such third phase for future \GE\ detectors is considered beyond the scope of this work, focusing on experimental proposals for the coming decade.}.
 
\textbf{CUORE} \cite{Arnaboldi:2002du, Sisti:2010zz, Bellini:2009zw} is an array of TeO$_2$ bolometers. Because $^{130}$Te has a large natural abundance ($\sim$ 34\%), the need for enrichment is less important than in other isotopes.  CUORE can collect a large mass of isotope ($\sim 200$~kg for a total detector mass of about 740 kg). The advantages of the technique are similar to those of germanium experiments with about the same energy resolution and efficiency for the signal. Studies have already demonstrated background rates at the 0.2 \ckky\ level. We assign $4\times10^{-2}\ \ckky$ and $10^{-3}\ \ckky$ to the reference and optimistic scenarios respectively. Concerning the mass, we assume the full 200 kg of CUORE in the R scenario, and a possible doubling (by enriching the target to some 70\%) of the mass for the O scenario.  

The second approach considered is \textbf{xenon time-projection chambers}.
Xenon is a suitable detection medium, providing both scintillation and ionization signals. It has a \bb-decaying isotope, \XE, with a natural abundance of about 10\%. Compared to other \bb\ sources, xenon is easy (thus relatively cheap) to enrich in the candidate isotope. It has no other long-lived radioactive isotopes, and no spallation products constitute a background. There are two possibilities for a xenon TPC: a cryogenic liquid xenon TPC (LXe), or a (high pressure) gas chamber (HPXe). We consider examples of both options here.

The Enriched Xenon Observatory (\textbf{EXO}) \cite{Ackerman:2009br, Gornea:2004rg} will search for \bbonu\ decay in $^{136}$Xe using a 200-kg liquid-xenon (enriched at 80\% in \XE) TPC during its first phase. The use of liquefied xenon results in a relatively modest energy resolution, 3.3\% FWHM at \Qbb\ \cite{Conti:2003av} (using the anti-correlation between ionization and scintillation). The signal detection efficiency is estimated by the Collaboration to be about 70\%. Background rates of order $10^{-3}$ \ckky\ are expected in EXO-200. The improvement with respect to the high-resolution calorimeters comes from the event topological information, that allows the rejection of superficial backgrounds and multi-hit events, so that only energetic gammas from \TL\ and \BI\ constitute a significant source of background. The optimistic scenario, $5\times10^{-4}\ \ckky$, projects a further improvement in the detector radiopurity.  The ultimate goal of the EXO Collaboration is to develop the so-called \emph{barium tagging} \cite{Danilov:2000pp}. This technique would allow the detection of the ion product of the $^{136}$Xe decay, and thus eliminate all backgrounds but the intrinsic \bbtnu. For this study, however, we have not considered its benefits.   Concerning the mass, we take 200 kg of Xe (160 kg of \XE) for the reference scenario, and assume one ton of isotope for the optimistic scenario.

The \textbf{NEXT} Collaboration is building a 100 kg high-pressure gaseous xenon (enriched at 90\% in \XE) TPC \cite{Granena:2009it}. The experiment aims to take advantage of both good energy resolution ($\lesssim 1\%$ FWHM at \Qbb ) and the presence of a \bbonu\ topological signature for further background suppression. As a result, the background rate is expected to be one of the lowest among all the proposals considered here: $2\times10^{-4}$ \ckky\ in the reference scenario, with a further improvement by a factor of 3 in the optimistic scenario. The low background rate, however, comes at the expense of a relatively inefficient signal detection: $\sim30\%$. In order to reach its ambitious goals, the NEXT Collaboration plans to rely on electroluminescence to amplify the ionization signal, using two separate photo-detection schemes for an optimal measurement of both calorimetry and tracking \cite{Nygren:2009zz}. Like in the case of EXO we assume 100 kg of xenon (90 kg of \bb\ isotope) for the reference scenario and 1 ton for the optimistic scenario.

The third category of experiments is \textbf{large self-shielding calorimetry}. These are large detectors in which the source, dissolved in liquid scintillator, is surrounded by a large buffer volume. The advantages of the approach are the self-shielding against external backgrounds and the scalability, since it appears possible to dissolve large isotope masses. The main disadvantage is poor energy resolution. 

\textbf{SNO+} \cite{Chen:2008un, Wright:2009, Kraus:2010zzb} proposes to fill the Sudbury Neutrino Observatory (SNO) with ultra-pure liquid scintillator. A mass of several tens of kilograms of \bb-decaying material can be added to the experiment by dissolving a neodymium salt in the scintillator. The natural abundance in the \ND\ isotope is 5.6\%. Given the liquid scintillator light yield and photocathode coverage of the experiment, a modest energy resolution performance (about 6\% FWHM at \Qbb) is expected. External backgrounds can be rejected with a relatively tight fiducial volume selection, cutting about $\sim50\%$ of the signal. A background rate similar to that of GERDA in its first phase is expected ($\sim$10$^{-2}$ \ckky\ in the R scenario, improving by a factor 10 in the O scenario. Concerning the isotope mass, SNO+ has the disadvantage of using natural neodymium, resulting in about 50 kg of isotope for the reference scenario. The optimistic scenario assumes that enriched neodymium can be use, resulting in a factor 10 increase of the mass. 

The \textbf{KamLAND-Zen} \cite{Terashima:2008zz, Koga:2010ic} experiment plans to dissolve 400 kg of \XE\ in the liquid scintillator of KamLAND in the first phase of the experiment, and up to 1 ton in a projected second phase. Xenon is relatively easy to dissolve (with a mass fraction of more than 3\% being possible) and also easy to extract. The major modification to the existing KamLAND experiment is the construction of an inner, very radiopure ($10^{-12}$ g/g of $^{238}$U and $^{232}$Th) and very transparent balloon to hold the dissolved xenon. The balloon, 1.7 meters in radius, would be shielded from external backgrounds by a large, very radio-pure liquid scintillator volume. While the energy resolution at \Qbb\ (about 10\%) is inferior to that of SNO+, the detection efficiency is much better ($80\%$) due to its double envelope. The double envelope design is also responsible for the low expected background rate, $2\times10^{-4}$ \ckky\ in the R scenario, with a further improvement by a factor of 2 in the O scenario.

The fourth and last category is that of \textbf{Tracko-Calo experiments}, where foils of \bb\ source are surrounded by a tracking detector that provides a direct detection of the two electron tracks emitted in the decay. The quintessential example of this technique is the SuperNEMO experiment. 

\textbf{SuperNEMO} \cite{Saakyan:2009zz, Shitov:2010nt, Novella:2008} is a series of modules, each one consisting of a tracker and a calorimeter that surround a thin foil of the \bb\ isotope. In SuperNEMO the target will likely be \SE, although other isotopes such as \ND\ or $^{48}$Ca are also being considered. The mass of the target is limited to a few kg (typically 5 to 7) by the need to build it foil-like, and to minimize multiple scattering and energy loss. The tracker and  calorimeter can record the trajectory of the charged particles and measure their energies independently. As shown by the successful NEMO-3 experiment \cite{Arnold:2005rz, Argyriades:2008pr, Argyriades:2009vq}, this technique, which exploits maximally the topological signature of the events, leads to excellent background rejection. Moreover, Tracko-Calo experiments allow for the determination of the \bbonu\ decay mechanism as the individual energies and trajectories of both electrons are measured \cite{Arnold:2010tu}. In exchange, the selection efficiency is --- like in the case of the NEXT experiment --- relatively low (about 30\%), and the resolution rather modest (4\% FWHM at \Qbb). This technique is very hard to extrapolate to large masses due to the size, complexity and cost of each module. For this reason, we consider that the target mass of the SuperNEMO collaboration (100 kg, or about 20 modules) will only be achieved in the optimistic scenario. For the reference scenario we consider 7 kg. This is the isotope mass for the \emph{demonstrator} module that the Collaboration expects to commission in 2013.
\subsection{Validity of background rate assumptions} \label{subsec:results_b}
\begin{figure}[t!]
\centering
\includegraphics[width=0.90\textwidth]{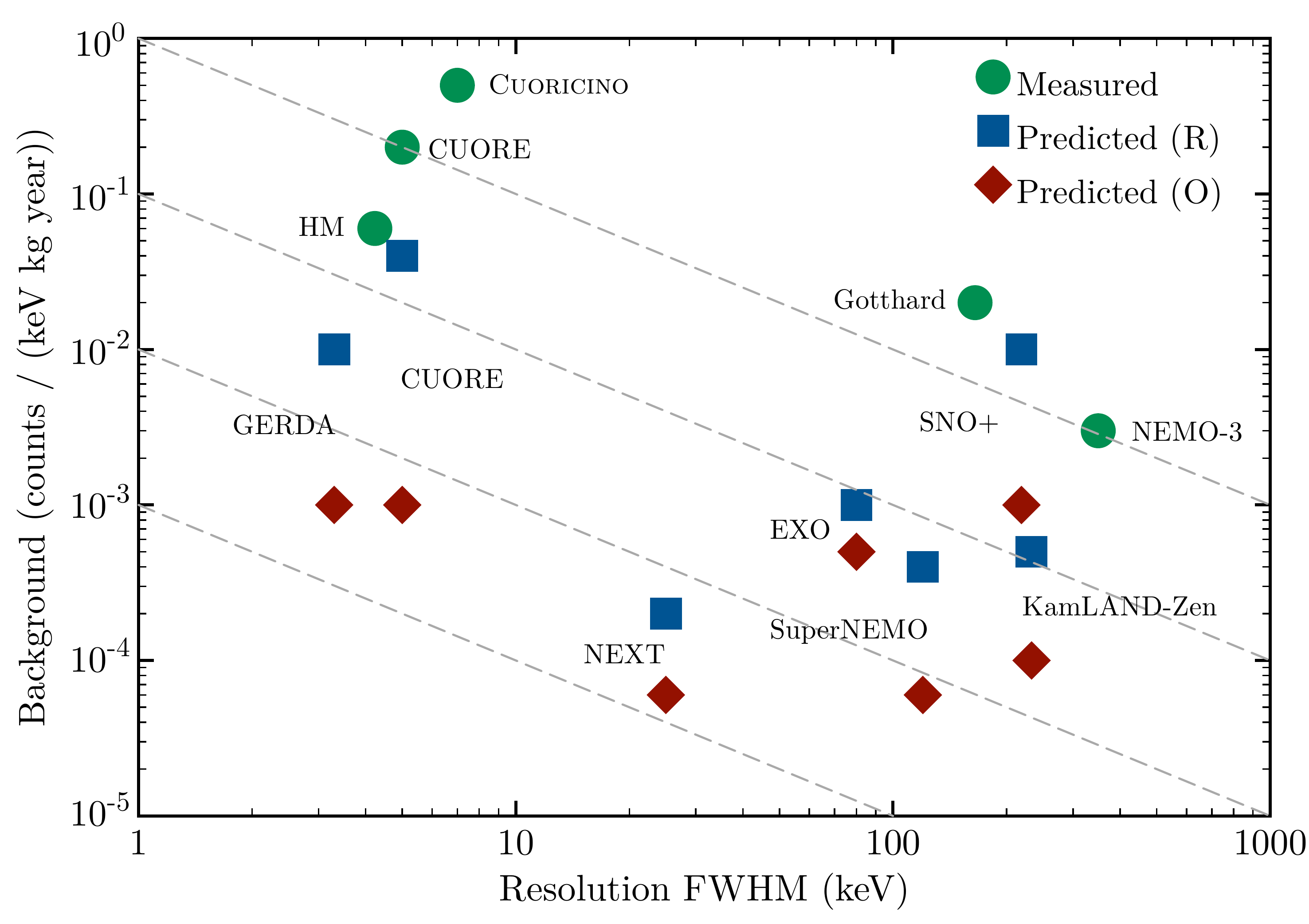}
\caption{Background rate in the ROI plotted versus the energy resolution (FWHM) for different past and present experiments. The dashed lines delimit bands where the experiments falling inside have background rates per unit of exposure --- counts/(kg$\cdot$year) --- of the same order of magnitude. (Figure adapted from \cite{Giuliani:2010ag}.)}
\label{fig.b_DE}
\end{figure}

How justified are the assumptions of the different experiments regarding their expected level of background suppression? Figure \ref{fig.b_DE} offers some clues. It shows the background rate in the ROI --- in \ckky\ --- versus the energy resolution (FWHM) of several past and present experiments. The (green) circles correspond to measured data, while the (blue) squares and (red) diamonds correspond, respectively, to the R and O background assumptions of the experiments discussed in the previous section. The dashed lines delimit bands where the experiments falling inside have background rates per unit of exposure --- counts/(kg$\cdot$y) --- of the same order of magnitude.

The best background rates obtained so far are rather modest if one considers the ambitions of the collaborations. Even in the case that we call \emph{reference}, achieving the expected performance poses a major experimental challenge.

The GERDA experiment expects to reduce the background rate achieved in Heidelberg-Moscow --- 0.06 \ckky\ after pulse-shape discrimination --- by at least a factor of 6 (phase I), and up to a factor of 60 (phase II). 
This improvement comes from a better shielding and improved radiopurity, and from active background suppression techniques such as rejection of multi-site depositions \cite{Barbeau:2007qi, Budjas:2009zu, Agostini:2010rd}.

CUORE aims at achieving a background rate of about $4\times10^{-2}$ \ckky. A previous effort, the \textsc{Cuoricino} detector \cite{Arnaboldi:2008ds}, a tower of bolometers that accumulated an exposure of 11.83 kg~$\cdot$~years, measured a background rate of 0.6 \ckky. More recent tests within the CUORE Collaboration have measured about 0.2 \ckky\ \cite{Sisti:2010zz}. Further improvements in the background rate would require, probably, active suppression methods such as those proposed by the LUCIFER project \cite{Giuliani:2010}. Such techniques may permit a reduction of the background rate down to $10^{-3}$ \ckky\ or better.

The xenon time-projection chambers have a precedent in the Gotthard experiment \cite{Luscher:1998sd}, a small xenon TPC ($\sim$5 kg) operated at 5 bar that obtained a background rate in the ROI of about 0.02 \ckky\ using event topological information to reject backgrounds. The NEXT experiment expects a factor of 100 improvement over Gotthard due to an upgraded detection technique: the Gotthard TPC suffered of modest energy resolution --- 6.6\% FWHM at the $Q$-value --- probably due to the addition of methane to the xenon (in order to increase the drift velocity and to suppress diffusion), which quenched the scintillation and ionization signals. NEXT aims at achieving energy resolution better than 1\% FWHM at \Qbb, enough to separate the \bbonu\ signal peak from the Compton spectrum of \TL, the main background source in Gotthard. Additionally, NEXT will measure the $t_{0}$ of each event using the primary scintillation signal, a handle that Gotthard lacked. This permits the definition of  a fiducial volume that eliminates any charged track emanating from detector surfaces. EXO projects a background level of $10^{-3}$ \ckky, thanks to their good self-shielding. This background rate is a factor 5 higher than that of NEXT and compatible with the availability of a topological signature in the pressurized gas, not present in the liquid phase.

SNO+ projects a background level similar to that of EXO, also due to good self-shielding. KamLAND-Zen, however, expects a factor of 5 lower rate than both EXO and SNO+ and at the same level of NEXT and SuperNEMO, in spite of lacking a topological signature. The advantage over SNO+ is the inner balloon that allows them to dissolve the xenon in a smaller volume of liquid scintillator. This reduces the fractional background due to the scintillator itself and improves the shielding. 

The NEMO-3 experiment has achieved a spectacular background rate of $3\times10^{-3}$ \ckky\ using the topological signature of the events to discriminate signal from background. SuperNEMO foresees an improvement of up to a factor of 60 in the background rate with respect to its predecessor. This implies challenging requirements in the radiopurity of the source foil and in the level of radon inside the detector (the two main sources of background in NEMO-3), since the experimental technique is essentially the same.

\section{Sensitivity of the proposals}\label{sec:results}
We now turn to the results concerning the \mbb\ sensitivity, as defined in Sec.~\ref{sec:stats}, for the various \bbonu\ proposals discussed in Sec.~\ref{sec:proposals}. In Sec.~\ref{subsec:results_full}, we show our results for the \mbb\ sensitivity of the proposals as a function of exposure, for both our reference and optimistic scenarios. Section~\ref{subsec:results_nextg} illustrates how the sensitivity of the proposals is affected by changes in the background rate assumptions. We comment on the scalability to large isotope masses of the various proposals, and on their plausible \mbb\ sensitivity after a common data-taking period of 10 years, in Sec.~\ref{subsec:scalability}. The efficiency of each experiment, as well as the overall efficiency for a ROI of 1 FWHM, is included in all the results.

A caveat is in order. Although we believe that our analysis is robust and simple, it also has some limitations. On the one hand, the quoted sensitivities could still improve somewhat if an energy-dependent analysis is performed. However, this requires a detailed understanding of the energy distribution of the various backgrounds in each experiment, and such information is hard to find in the existing literature. On the other hand, adding systematic errors in both the estimation of the background and the calculation of the efficiency will partially spoil the sensitivity. This last point may severely affect the sensitivity results, but it is very difficult to implement in a general study like ours, since the published information concerning systematic errors of the different proposals is very scarce.

\subsection{Sensitivity as a function of the exposure} \label{subsec:results_full}

\begin{figure}[t!b!]
\centering
\includegraphics[width=0.475\textwidth]{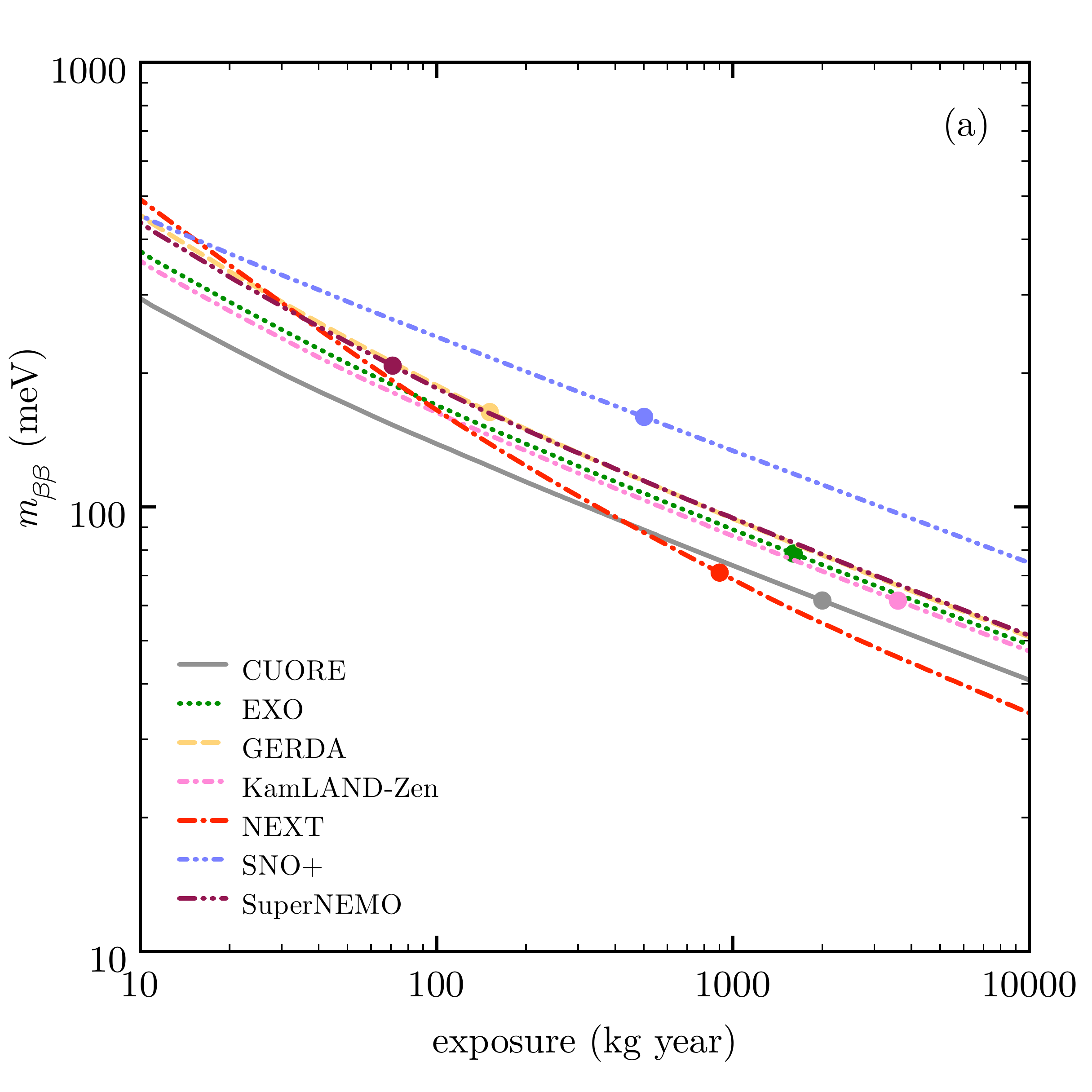}
\includegraphics[width=0.475\textwidth]{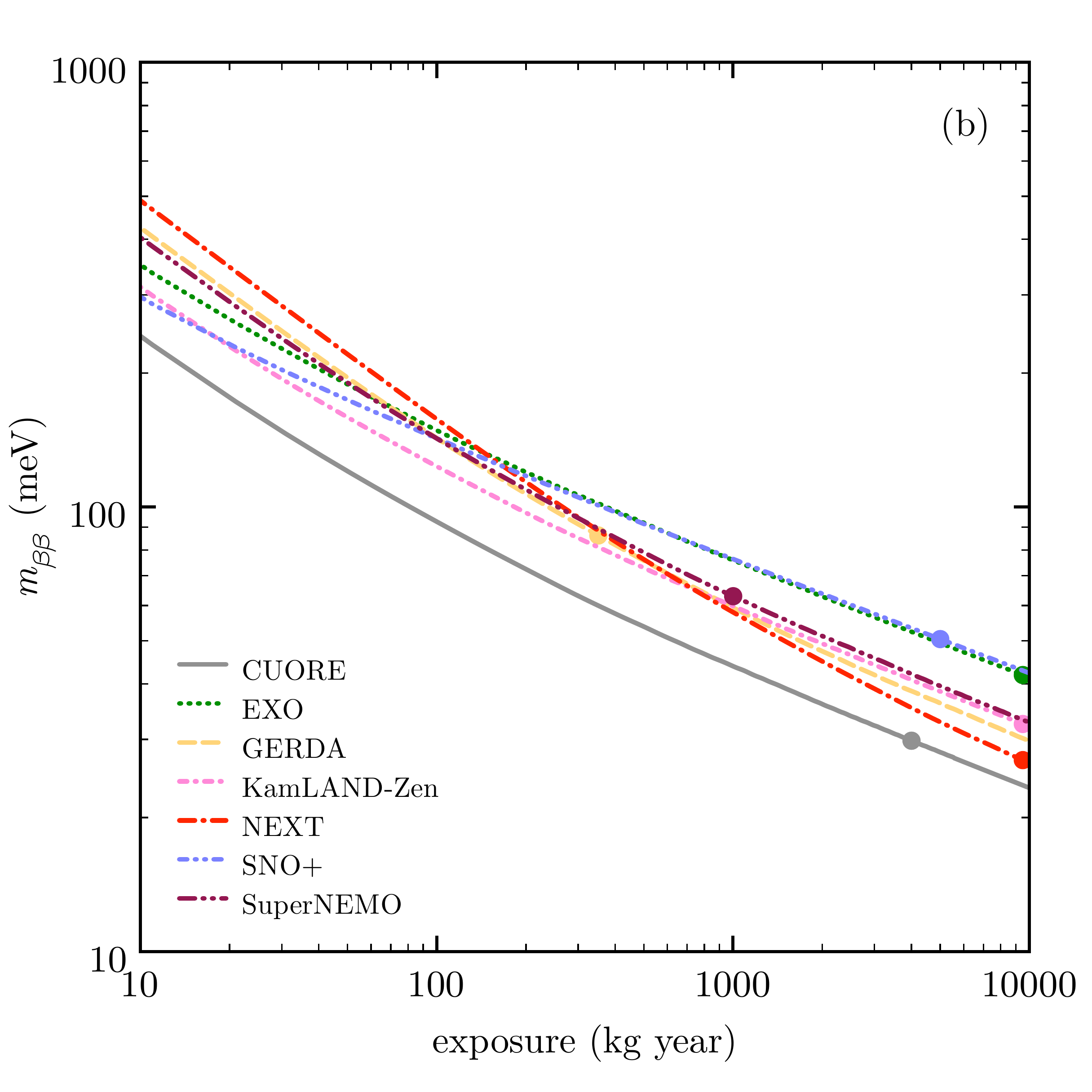}
\caption{The \mbb\ sensitivity (at 90\% CL) as a function of exposure of the seven different \bbonu\ proposals considered. For each proposal, two detector performance scenarios are shown: (a) reference case, (b) optimistic case. For illustrative purposes, the filled circles indicate 10 years of run-time according to the reference and optimistic isotope mass assumptions in Table \ref{tab:proposals}.} \label{fig.SensiF}
\end{figure}

Figure \ref{fig.SensiF} shows the \mbb\ sensitivity (at 90\% CL) as a function of exposure for the seven \bbonu\ proposals reviewed. We consider here not only the impact of isotope choice, efficiency and energy resolution, but also the effect of the expected background rate in the R and O scenarios. For the sake of simplicity, let us focus here on a fixed exposure of $10^3\ \kgy$ for all proposals.

In the reference scenario and for the above-mentioned exposure, NEXT and CUORE have the best sensitivities, reaching 69 and 74 meV at 90\% CL, respectively. KamLAND-Zen, EXO, GERDA and SuperNEMO follow, with sensitivities around 90 meV. SNO+ reaches a sensitivity 134 meV. Taking into account that these numbers represent a limit of maximum exposure for most of the proposals, 1 ton$\cdot$year, it follows that the ``next-generation'' experiments will, at best, explore the degenerate spectrum.  
 
In the optimistic scenario, the lower background regime for all experiments allows to obtain significantly better sensitivities for the same $10^3\ \kgy$ exposure. In this case, the best sensitivity is obtained by CUORE (44 meV at 90\% CL). NEXT, KamLAND-Zen, GERDA and SuperNEMO follow, with sensitivities around 60 meV. EXO and SNO+ reach sensitivities around 75 meV.

\subsection{Sensitivity as a function of background rate}\label{subsec:results_nextg}

Table \ref{tab.sensi} gives the sensitivity of all the proposals (for an exposure of 1 ton$\cdot$year) and for five different background rate assumptions. We can observe the following trends:
\begin{itemize}
\item
 In the background rate characterizing past \bbonu\ experiments, $c\sim 10^{-1}\ \ckky$, CUORE would be the experiment with best sensitivity, 93 meV at 90\% CL.
 \item
 With a one order of magnitude better background rate suppression, $c=10^{-2}\ \ckky$, GERDA would reach 91 meV at 90\% CL and CUORE would almost fully explore the degenerate spectrum (down to 54 meV at 90\% CL). 
 \item
 For background rates at the $10^{-3}\ \ckky$ level, CUORE would partially probe the inverted hierarchy (\mbb\ $\sim 34$ meV), GERDA would almost fully explore the degenerate spectrum, while EXO, NEXT and SNO+, would have a sensitivity better than 100 meV. 
 \item
 For even lower background rates, $c=10^{-4}\ \ckky$, NEXT, EXO, KamLAND-Zen and SuperNEMO would almost fully cover the degenerate spectrum; GERDA would probe a small part of the inverse hierarchy; and CUORE would cover it almost completely. 
 \item
 In the very low background rate limit of $c=10^{-5}\ \ckky$, one can see that the improvement in sensitivity of both CUORE and GERDA is very small, indicating that these experiments have reached the ``background-free'' regime for the considered exposure. NEXT, EXO, KamLAND-Zen and SuperNEMO, instead, keep improving their sensitivity, and explore partially the inverse hierarchy.
 \end{itemize}

\begin{table}[t!]
\begin{center}
\begin{tabular}{l D{.}{.}{-1} D{.}{.}{-1} D{.}{.}{-1} D{.}{.}{-1} D{.}{.}{-1}}
\hline\hline \noalign{\smallskip}
\multirow{2}{*}{Experiment} & \multicolumn{5}{c}{ \mbb\ (meV)} \\ \cline{2-6} \noalign{\smallskip}
 & \multicolumn{1}{c}{$c=10^{-1}$} & \multicolumn{1}{c}{$c=10^{-2}$} & \multicolumn{1}{c}{$c=10^{-3}$} & \multicolumn{1}{c}{$c=10^{-4}$} & \multicolumn{1}{c}{$c=10^{-5}$}\\ \hline \noalign{\smallskip}
CUORE          	&  93 &  54 &  34 & 25 & 24 \\
EXO            	& 258 & 147 &  85 & 51 & 37 \\
GERDA 			& 155 &  91 &  58 & 45 & 42 \\
KamLAND-Zen    	& 314 & 178 & 101 & 60 & 39 \\
NEXT           	& 269 & 154 &  91 & 61 & 50 \\
SNO+           	& 237 & 134 &  76 & 45 & 30 \\
SuperNEMO      	& 354 & 201 & 115 & 69 & 48 \\
\hline\hline
\end{tabular}
\end{center}
\caption{Large exposure \mbb\ sensitivity (for $Mt= 10^{3}\ \kgy$ and at 90\% CL) of the different \bbonu\ experiments in terms of the background rate (expressed in \ckky).}
\label{tab.sensi}
\end{table}%

\subsection{Scalability to large isotope masses}\label{subsec:scalability} 

Having considered \mbb\ sensitivity as a function of exposure in \kgy , we consider now the mass scalability of the different proposals. 

Germanium experiments such as GERDA and MAJORANA are easier to scale, given their compactness. While masses of the order of 100 kg appear possible, it may be difficult, in particular for economical reasons, to go much beyond that.
CUORE aims already at a large mass, and can, presumably, scale up at least by a factor two if enrichment at large scale is feasible and the backgrounds can be kept under control.

It will be very difficult to scale up SuperNEMO. One is faced here with the difficulty of replicating modules (about 20 are needed), high cost, and the need of a very large underground volume to host the modules and their shielding. The technique is clearly  unsuitable for large masses, and reaching the collaboration's target mass of 100 kg
appears as a formidable challenge.

The scalability of SNO+ depends on the feasibility of enriching neodymium, most likely a difficult enterprise. In case of success, they could multiply their mass by a factor 10.

All the xenon-based experiments can, in principle, use one ton of enriched xenon or more for the next-to-next generation.  Notice that 700 kg of \XE\ have already been acquired by the 3 collaborations. Clearly the best isotope among the ones considered here, as far as  scalability to large \bb\ isotope masses is concerned, appears to be \XE. 

\begin{table}[t!]
\begin{center}
\begin{tabular}{l c D{.}{.}{3.0} D{.}{.}{3.0} D{.}{.}{3.0} D{.}{.}{3.0} D{.}{.}{3.0} D{.}{.}{3.0}}
\hline\hline \noalign{\smallskip}
\multirow{2}{*}{Experiment} &  & \multicolumn{6}{c}{ \mbb\ (meV)} \\ \cline{3-8} \noalign{\smallskip}
 &  & \multicolumn{1}{c}{{\bf PMR}} & \multicolumn{1}{c}{GCM} & \multicolumn{1}{c}{IBM} & \multicolumn{1}{c}{ISM} & \multicolumn{1}{c}{QRPA(J)} & \multicolumn{1}{c}{QRPA(T)} \\ \hline \noalign{\smallskip}
\multirow{2}{*}{CUORE}			& R & {\bf 62} & 44 & 47 & 84 & 44 & 66 \\
								& O	& {\bf 30} & 21 & 23 & 41 & 21 & 32 \\ \hline \noalign{\smallskip}
\multirow{2}{*}{EXO} 	       	& R & {\bf 79} & 53 & 56 & 101 & 66 & 96 \\
								& O & {\bf 41} & 28 & 29 &  53 & 35 & 51 \\ \hline \noalign{\smallskip}
\multirow{2}{*}{GERDA}	 		& R & {\bf 163} & 144 & 103 & 204 & 120 & 157 \\
								& O &  {\bf 86} &  76 &  54 & 108 &  64 &  83 \\ \hline \noalign{\smallskip}
\multirow{2}{*}{KamLAND-Zen} 	& R & {\bf 62} & 41 & 44 & 79 & 52 & 75 \\
								& O & {\bf 32} & 22 & 23 & 41 & 27 & 39 \\ \hline \noalign{\smallskip}
\multirow{2}{*}{NEXT}        	& R & {\bf 71} & 48 & 51 & 92 & 60 & 87 \\
								& O & {\bf 27} & 18 & 19 & 34 & 22 & 33 \\ \hline \noalign{\smallskip}
\multirow{2}{*}{SNO+}        	& R & {\bf 159} & 217 & 136 & - & - & 118 \\
								& O &  {\bf 50} &  69 &  43 & - & - &  37 \\ \hline \noalign{\smallskip}
\multirow{2}{*}{SuperNEMO}   	& R & {\bf 208} & 171 & 139 & 274 & 158 & 174 \\
								& O &  {\bf 63} &  52 &  42 &  83 &  48 &  53 \\ 
\hline\hline
\end{tabular}
\end{center}
\caption{Sensitivity of the different proposals at 90\% CL in the R and O case, taking into account also the projected isotope mass that can be collected by the different proposals and for a 10 years data-taking period. All sensitivity numbers refer to \mbb\ values in meV units. The PMR column relies on our estimate for NME most probable values, while subsequent columns use published NME results (see Sec.~\ref{sec:bb} for details).}
\label{tab.sscale}
\end{table}%

Table \ref{tab.sscale} summarizes our \mbb\ sensitivity findings if we further assume the reference and optimistic values for the \bb\ isotope masses as given in Tab.~\ref{tab:proposals}, and for a data-taking period of 10 years, common to all proposals. In the reference scenario, CUORE, EXO, KamLAND-Zen and NEXT cover most of the degenerate spectrum, with CUORE and KamLAND-Zen reaching the best sensitivity ($\mbb\simeq 62$ meV at 90\% CL assuming our PMR estimate for NME most probable values). In the optimistic scenario, these four experiments cover part of the inverted hierarchy, with NEXT reaching the best sensitivity ($\mbb\simeq 27$ meV at 90\% CL). Even taking into account the uncertainties associated to this calculation, it appears that the scalability to large masses and expected low background of the xenon experiments give them an advantage over other experimental approaches, with the possible exception of CUORE. 

For completeness, Tab.~\ref{tab.sscale} also shows how \mbb\ sensitivity results are affected by different assumptions regarding NME values. When available, ISM NME values \cite{Menendez:2009di, Menendez:2009oc} give \mbb\ sensitivity results that are about 30\% worse than what obtained from our estimate for NME most probable values (PMR column in Tab.~\ref{tab.sscale}). By construction, non-ISM NME values \cite{Rodriguez:2010mn, Barea:2009zza, Suhonen:2010ci, Simkovic:2008fr, Simkovic:2009cu,Simkovic:2009fv,Fang:2010qh} tend to give \mbb\ sensitivity results that are better than the PMR ones by a similar relative amount.

\section{Conclusions}\label{sec:conclusions}
The answer to the question of the neutrino particle/antiparticle nature has become a critical issue after the observation of neutrino oscillations. 
Not only it would improve our understanding of this intriguing fermion, but it would also have an enormous impact in many other areas of fundamental physics.

Neutrinoless double beta decay (\bbonu) is possible if and only if neutrinos are massive Majorana particles. 
The observation of such a process is considered the most promising
strategy to demonstrate the Majorana nature of the neutrino. Furthermore, the measurement
of the lifetime for this process would provide direct information on the absolute scale of 
neutrino masses.

An obvious goal for the next-generation experiments is to push the current limits down to Majorana masses \mbb\ of at least 100 meV to unambiguously confirm or discard the evidence for \bbonu\ claimed by part of the Heidelberg-Moscow Collaboration \cite{KlapdorKleingrothaus:2001ke}, with sufficient statistics. This appears to be within the range of all the proposals. A more ambitious goal is to fully explore the degenerate spectrum, down to $\mbb\simeq$ 50 meV. None of the proposals appears quite capable of this, under the assumptions discussed in this work (the reference scenario, 90\% confidence level, data-taking period of 10 years), although CUORE ($\mbb\simeq$ 62 meV), KamLAND-Zen ($\mbb\simeq$ 62 meV), NEXT ($\mbb\simeq$ 71 meV) and EXO ($\mbb\simeq$ 79 meV) end up quite close. 

The optimistic scenario described in the text may be implemented by the next-to-next generation of experiments. The goal here would
be to at least partially explore the inverse hierarchy. This seems to be within the reach of NEXT ($\mbb\simeq$ 27 meV),
CUORE ($\mbb\simeq$ 30 meV), KamLAND-Zen ($\mbb\simeq$ 32 meV) and perhaps EXO ($\mbb\simeq$ 41 meV).

More generally, the suitability of SNO+, GERDA and SuperNEMO for a next-to-next round of experiments is essentially dependent upon their capability to scale the technology to large masses. Scalability appears easier for CUORE, in particular if the techniques being developed in the context of the LUCIFER Collaboration (see for example \cite{Giuliani:2010}) permit large background rate reductions. Finally it appears that \XE\ would be a a particularly favorable isotope to use,  since it permits target masses of 1 ton or more and low-background experimental techniques such as the ones proposed by NEXT, KamLAND-Zen and EXO. 

Last but not least, while we believe that the general trends observed in the sensitivity comparisons make sense, one should take the numbers presented here \textit{cum grano salis}. There are many uncertainties, ranging from the precise values of the nuclear matrix elements to the ultimate background rate and isotope mass that can be achieved by the different experimental techniques. Nevertheless, we suggest that \bbonu\ collaborations adopt the simple and unambiguous procedure we have described in this work to derive \mbb\ sensitivities, in order to allow for more meaningful physics case comparisons.

\acknowledgments
The authors acknowledge support by the Spanish Ministry of Science and Innovation (MICINN) under grants CONSOLIDER-INGENIO 2010 CSD2008-0037 (CUP), FPA2009-13697-C04-04 and RYC-2008-03169. The authors thank Bob Cousins, Fred James, Pilar Hern\'andez and Mauro Mezzetto for illuminating discussions. We also appreciate the comments of several of our colleagues in the double beta decay community, including Alessandro Bettini, Peter Grabmayr, Franco Iachello, Stefan Sch\"onert, Steve Elliott, Giorgio Gratta, and Andrea Giuliani. 

\appendix
\section{Construction of unified-approach confidence intervals}\label{sec:fc}
For completeness and for pedagogical purposes, in this appendix we describe in some detail how to construct a confidence interval for a signal mean to be inferred from observations, using the unified approach by Feldman and Cousins \cite{Feldman:1997qc} and in the case of a Poisson process with background. In this case, the relevant pdf to observe $n$ events given a (unknown) signal mean $\mu$ and a (known) background mean $b$ is given by the Poisson distribution:
\begin{equation}
{\rm Po}(n;\mu+b) = \frac{(\mu+b)^n}{n!} \ e^{-(\mu+b)},
\end{equation}

Just as can be done for classical confidence intervals (that is, the one-sided upper confidence limit and the two-sided central confidence interval discussed in Secs.~\ref{subsec:stats_nobgr} and \ref{subsec:stats_bgr}), the unified approach uses the Neyman construction of confidence belts \cite{Neyman:1937}. In this construction, given the known mean background expectation $b$ and for each value of the unknown signal mean $\mu$, one selects an interval in $n$ such that:
\begin{equation}
\sum_{n=n_1}^{n_2}{\rm Po}(n;\mu+b) \ge {\rm CL}
\label{eq.acceptance_interval}
\end{equation}
where ${\rm CL}$ is the desired confidence level, for example ${\rm CL}=0.90$. Once this exercise is repeated for all possible values of $\mu$ and once the experiment's outcome $n_{\rm obs}$ is known, the confidence interval $[\mu_{\rm lo},\mu_{\rm up}]$ for the signal mean $\mu$ can be extracted. 

However, while the acceptance interval $n \in [n_1,n_2]$ is constrained by Eq.~\ref{eq.acceptance_interval}, it is not fully specified by it. The freedom on how to specify such intervals marks the difference between the classical and the unified approach confidence belt construction. In the unified approach method, such acceptance intervals at fixed $\mu$ are determined based on a ordering principle for the $n$ values based on likelihood ratios, as will be shown below. The entire procedure can be summarized as follows:
\begin{enumerate}
\item Compute the mean background expectation, $b$. Suppose that, in our case, $b=1.0$. Note that $b$ does not need to be an integer value, in general. 
\item For the given mean background expectation $b$ and for each possible measurement outcome $n$, compute the best estimator $\mu_{\rm best}$ for the true value of the mean signal yield $\mu$. If $\mu$ were unconstrained, the best estimator $\mu_{best}$ would simply be found by maximizing the Poisson probability for any given $b$ and $n$:
\begin{equation}
\left. {\frac{{{\rm dPo}(n;\mu+b)}}{{{\rm d}\mu }}} \right|_{n,b}  = 0 \Rightarrow \mu_{\rm best} =n-b.  
\end{equation}  
However, we know that only non-negative values for $\mu$ are physically allowed, so that our best estimator is given by:
\begin{equation}
\mu_{\rm best} =\max(0,n-b)
\end{equation}
In the equations above, the measurement outcome $n$ can take any non-negative integer value, while $\mu_{\rm best}$ can take any non-negative value. The values of $\mu_{\rm best}$ for $b=1.0$ and for each possible measurement outcome $n$ are reported in Tab.~\ref{t.mubest}.
\begin{table}[bht!]
\begin{center}
\begin{tabular}{c|c|c}
\hline\hline
$n$ & $\mu_{\rm best}$ & ${\rm Po}(n;\mu_{\rm best}+b)$ \\ \hline
0 & 0.0 & 0.368 \\
1 & 0.0 & 0.368 \\
2 & 1.0 & 0.271 \\
3 & 2.0 & 0.224 \\
4 & 3.0 & 0.195 \\
5 & 4.0 & 0.175 \\ \hline\hline
\end{tabular}
\end{center}
\caption{\label{t.mubest}For a given mean background expectation $b=1.0$, values for the signal best estimator $\mu_{\rm best}$ and for ${\rm Po}(n;\mu_{\rm best}+b)$ as a function of all possible measurement outcomes $n$.}
\end{table}
\begin{figure}[btp]
\begin{center}
\includegraphics[width=0.5\textwidth]{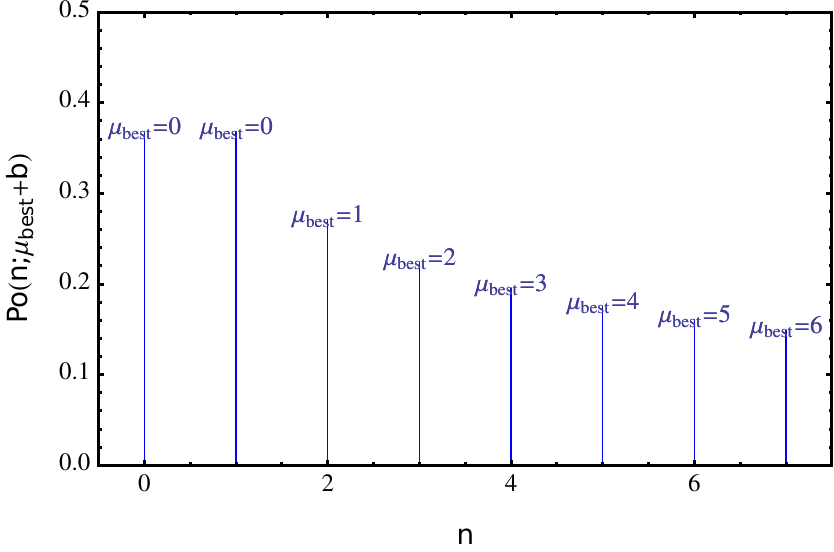}
\caption{\label{fig.mubest}For a given mean background expectation $b=1.0$, ${\rm Po}(n;\mu_{\rm best}+b)$ as a function of all possible measurement outcomes $n$ is shown. For each $n$, the value $\mu_{\rm best}$ is also shown in the figure,}
\end{center}
\end{figure}
\item For the given mean background expectation $b$ and for each possible measurement outcome $n$, compute the likelihood ${\rm Po}(n;\mu_{\rm best}+b)$ of obtaining $n$ given the best-fit physically allowed signal mean $\mu_{\rm best}$. This is reported in tabular form in Tab.~\ref{t.mubest}, and in graphical form in Fig.~\ref{fig.mubest}. Explicitly, the function ${\rm Po}(n;\mu_{\rm best}+b)$ is equal to:
\begin{equation}
{\rm Po}(n;\mu_{\rm best}+b) =
\begin{cases}
\frac{b^n}{n!}\ e^{-b}, & \mbox{if } n \le b\ (\Rightarrow \mu_{\rm best}=0) \\
\frac{n^n}{n!}\ e^{-n}, & \mbox{if } n > b\ (\Rightarrow \mu_{\rm best}>0)
\end{cases} 
\end{equation}
\item We now consider a possible value for the unknown true signal mean $\mu$. For simplicity, we will show the computational details only for $\mu =0$ and for $\mu = 1.0$. Note that $\mu$ does not need to be an integer value, in general.
\item For the given mean background expectation $b$ (fixed in step 1), and for the given true signal mean $\mu$ (fixed in step 4), order the possible measurement outcomes $n$ from most to least likely. This ordering is done according to the values of the likelihood ratio  (also called {\em profile likelihood}) \cite{Rolke:2004mj}: 
\begin{equation}
\mathcal{L}_R \equiv \frac{{\rm Po}(n;\mu+b)}{{\rm Po}(n;\mu_{\rm best}+b)},
\label{eq.lr}
\end{equation}
where the likelihood ${\rm Po}(n;\mu_{\rm best}+b)$ was computed in step 3, while ${\rm Po}(n;\mu+b)$ is the likelihood of obtaining $n$ for the given signal mean $\mu$. Explicitly, the function $\mathcal{L}_R$ is equal to:
\begin{equation}
\mathcal{L}_R =
\begin{cases}
(\frac{\mu+b}{b})^n \ e^{-\mu}, & \mbox{if } n \le b\ (\Rightarrow \mu_{\rm best}=0) \\
(\frac{\mu+b}{n})^n \ e^{-(\mu+b-n)}, & \mbox{if } n > b\ (\Rightarrow \mu_{\rm best}>0)
\end{cases} 
\end{equation}

 The highest rank ($\mathcal{R}=1$) is assigned to the $n$ value having the highest value of $\mathcal{L}_R$. 

As examples, ${\rm Po}(n;\mu+b)$ values for each $n$ are given in the left panels of Tab.~\ref{t.muexamples} and of Fig.~\ref{fig.Ponmu} for $\mu =0$, and in the right panels of Tab.~\ref{t.muexamples} and of Fig.~\ref{fig.Ponmu} for $\mu =1.0$. The corresponding likelihood ratios $\mathcal{L}_R$ and ranking of $n$ values from most to least likely are also given in Tab.~\ref{t.muexamples} for $\mu =0$ and 1.0. The likelihood ratio distributions are also shown in graphical form in Fig.~\ref{fig.LR}, again for $\mu=0$ (left) and $\mu =1.0$ (right). Note that $\mathcal{L}_R$ is monotonically decreasing with $n$ for $\mu=0$, but this is not the case for $\mu=1.0$, where $n=2$ has the maximum rank.
\begin{table}[bht!]
\begin{tabular}{c|c|c|c|c}
\hline\hline
$n$ & ${\rm Po}(n;\mu+b)$ & $\mathcal{L}_R$ & $\mathcal{R}$ & $\mathcal{P}$ \\ \hline 
0   & 0.368      & 1.000           & 1             & 0.368 \\
1   & 0.368      & 1.000           & 2             & 0.736 \\
2   & 0.184      & 0.680           & 3             & 0.920 \\
3   & 0.061      & 0.274           &               &       \\
4   & 0.015      & 0.078           &               &       \\
5   & 0.003      & 0.017           &               &       \\ \hline \hline
\end{tabular}
\hfill
\begin{tabular}{c|c|c|c|c}
\hline\hline
$n$ & ${\rm Po}(n;\mu+b)$ & $\mathcal{L}_R$ & $\mathcal{R}$ & $\mathcal{P}$   \\ \hline 
0   & 0.135      & 0.368           & 5             &  0.947 \\
1   & 0.271      & 0.736           & 3             &  0.722 \\
2   & 0.271      & 1.000           & 1             &  0.271 \\
3   & 0.180      & 0.805           & 2             &  0.451 \\
4   & 0.090      & 0.462           & 4             &  0.812 \\
5   & 0.036      & 0.206           &               &        \\ \hline\hline 
\end{tabular}
\caption{\label{t.muexamples}For a given mean background expectation $b=1.0$ and mean signal $\mu =0$ (left table) and $\mu =1.0$ (right table), ranking $\mathcal{R}$ of the $n$ values according to the likelihood ratio $\mathcal{L}_R={\rm Po}(n;\mu+b)/{\rm Po}(n;\mu_{\rm best}+b)$, and confidence interval construction in $n$ until a cumulative probability $\mathcal{P}=0.90$ is reached. See text for details.}
\end{table}
\begin{figure}[btp]
\begin{center}
\includegraphics[width=0.45\textwidth]{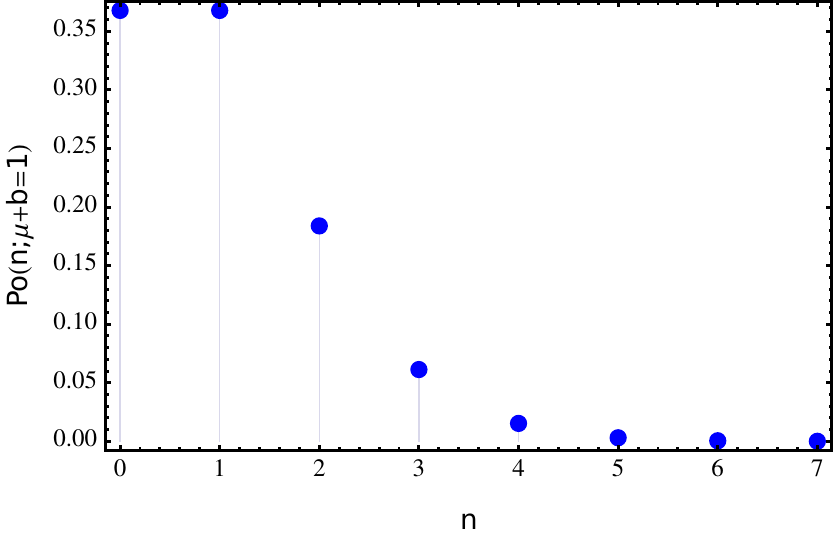}
\hspace*{1cm}
\includegraphics[width=0.45\textwidth]{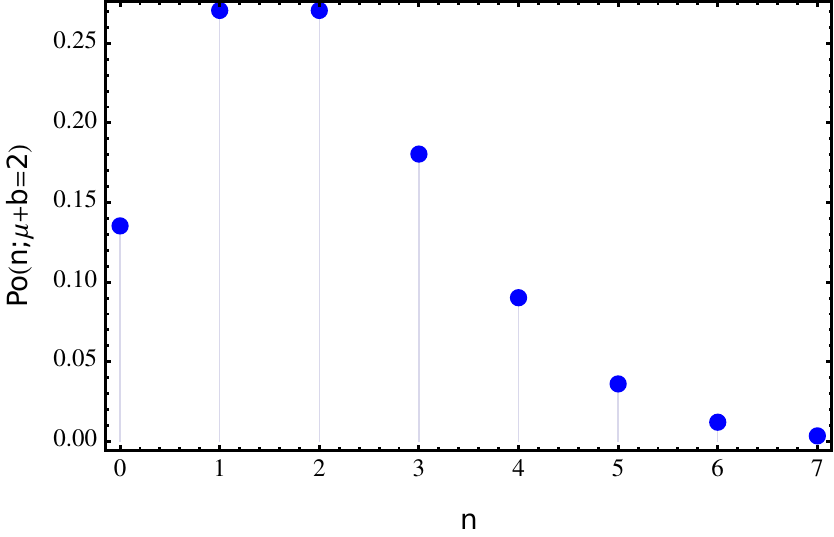}
\caption{The function ${\rm Po}(n;\mu+b)$ for a mean background expectation $b=1.0$, and a signal mean $\mu = 0$ (left) or $\mu = 1.0$ (right).}
\label{fig.Ponmu}
\end{center}
\end{figure}

\begin{figure}[btp]
\centering
\includegraphics[width=0.45\textwidth]{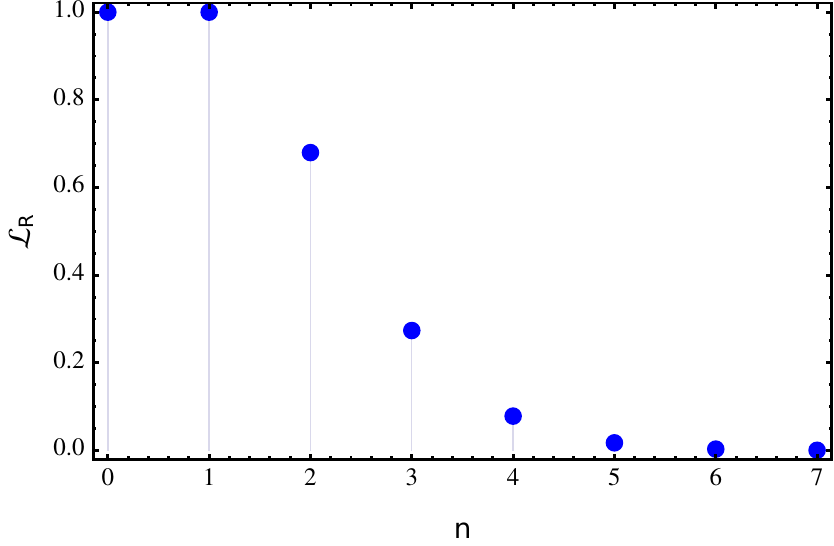}
\hspace*{1cm}
\includegraphics[width=0.45\textwidth]{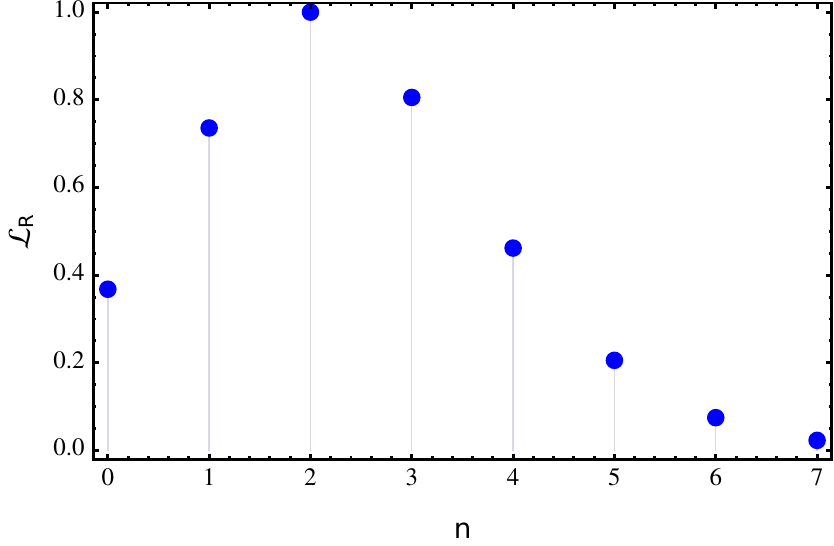}
\caption{The function  $\mathcal{L}_R = {\rm Po}(n;\mu+b)/{\rm Po}(n;\mu_{best}+b)$ for $\mu=0$ (left) and $\mu=1.0$ (right).}
\label{fig.LR}
\end{figure}
\item For a given $b$ and $\mu$, construct a confidence interval in $n$ by adding the ${\rm Po}(n;\mu+b)$ values until the cumulative probability is at least as large as the desired confidence level. In this confidence interval construction, the ${\rm Po}(n;\mu+b)$ values are added according to their ranking (as defined in step 5), starting from the most likely $n$ value ($\mathcal{R}=1$). As examples, Tab.~\ref{t.muexamples} (left) shows that only three $n$ values are necessary to reach a 90\% CL for the $\mu =0$ case. The three values are added in this order: $n=0,1,2$. The resulting confidence interval is therefore $n \in [0,2]$. Similarly, Tab.~\ref{t.muexamples} (right) shows that five $n$ values are needed to reach a 90\% CL: $n=2,3,1,4,0$, resulting in the interval $n \in [0,4]$. The larger the desired confidence level is, the larger is the corresponding $n$ interval.
\item Repeat steps 4-6 for different values of $\mu$. In this scan over all possible values of the unknown true signal mean $\mu$, one typically starts from $\mu =0$, and gradually increases the $\mu$ values according to some $\mu$ step size. Even though we showed the cases $\mu=0$ and $\mu=1.0$ only, the step size is often chosen to be much smaller than unity when small signals are searched for. 

 Once the $n$ confidence intervals have been obtained for each $\mu$, the confidence belt construction is complete for the known mean background expectation $b$ and for the desired confidence level (90\% CL, in our case). Note that this confidence belt construction does not require any knowledge of the actual experiment's outcome, $n_{\rm obs}$.
\item Now perform the measurement. Suppose that the result is $n_{\rm obs}=1$, that is, compatible with the background-only mean expectation $b=1.0$. The left panel in Fig.~\ref{fig.Belt} shows a vertical line corresponding to this value. The intercepts of the confidence belts with the $n_{\rm obs}=1$ vertical line fixes the allowed range in the true signal mean, $\mu \in [\mu_{\rm lo},\mu_{\rm up}]$. Given that the observable $n$ is discrete, one more prescription is needed in this case to fully specify the range $\mu \in [\mu_{\rm lo},\mu_{\rm up}]$. When several values of $\mu$ yield the same acceptance interval $n \in [n_1,n_2]$ in Eq.~\ref{eq.acceptance_interval}, the constant $n_{\rm obs}$ vertical line does not intersect the belts in only two points, but rather along two vertical segments. When this occurs, we conservatively take the confidence interval to have $\mu_{\rm lo}$ corresponding to the smallest value of $\mu$ with $n_2 = n_{\rm obs}$, and $\mu_{\rm up}$ as the largest value of $\mu$ with $n_1 = n_{\rm obs}$. As can be read from the left panel in Fig.~\ref{fig.Belt}, in this case the range is $\mu \in [0,3.4]$. On the other hand, suppose that the experiment had measured a significant excess above the background-only prediction: $n_{\rm obs}=10$. The right panel in Fig.~\ref{fig.Belt} shows the $\mu$ range that would have been obtained in this case: $\mu \in [4.5,15.5]$. 
\end{enumerate}

\begin{figure}[bht!]
\centering
\includegraphics[width=0.45\textwidth]{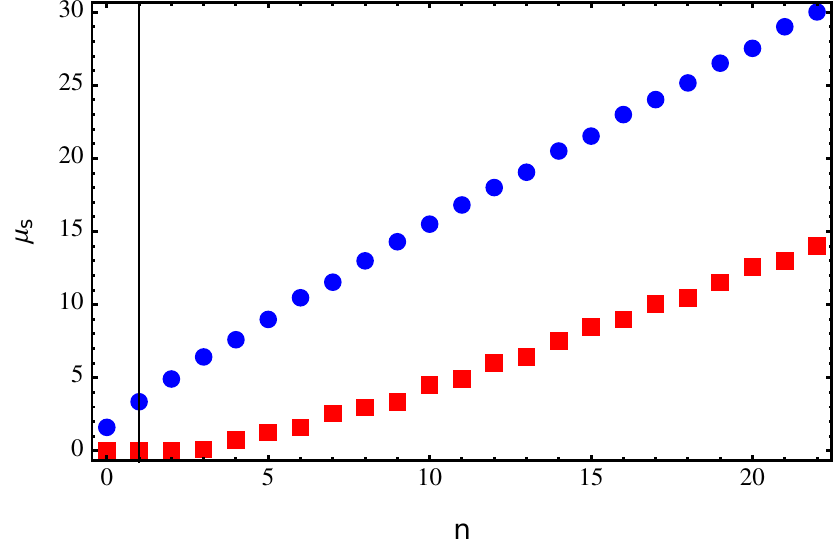}
\hspace*{1cm}
\includegraphics[width=0.45\textwidth]{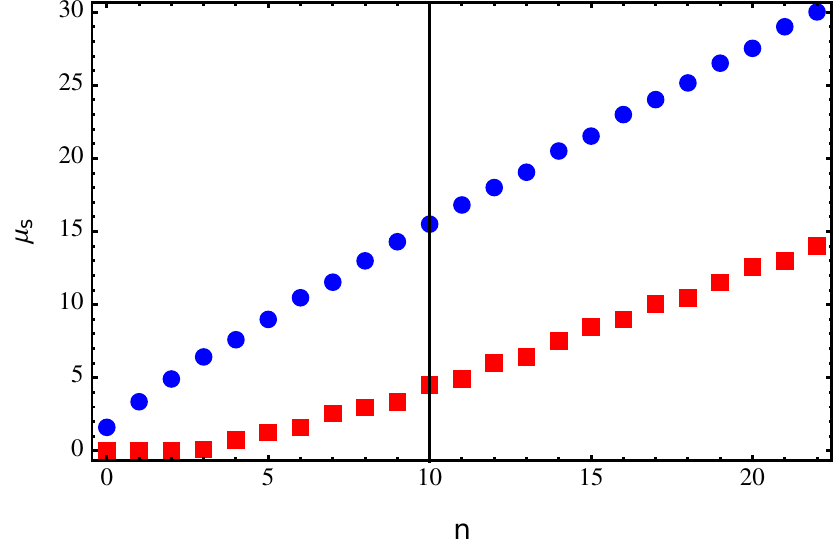}
\caption{\label{fig.Belt}Confidence belts constructed for a Poisson process with mean background expectation $b=1.0$, using the unified approach and for a 90\% confidence level. The left (right) panel shows how to infer a range in the signal yield $\mu$ given a measurement outcome of $n_{\rm obs}=1 (10)$.}
\end{figure}

The unified approach procedure described above solves two problems of classical confidence belts, namely that: (a) a downward fluctuation in background can produce an empty set confidence level (see Sec.~\ref{subsec:stats_bgr}), and (b) using the measured result to decide whether to use a central or upper ordering principle leads to the wrong coverage (\emph{flip-flopping}, see \cite{Feldman:1997qc}). The unified approach never produces empty confidence intervals, and provides the correct frequentist coverage by avoiding to use the measurement outcome to construct the confidence belts. In this approach, if the lower bound obtained by the above procedure is strictly $\mu_{\rm lo}=0$, then a upper limit is quoted; otherwise, if $\mu_{\rm lo}>0$, a central confidence interval is given. With these definitions of upper limit and central interval, and as can be seen in Fig.~\ref{fig.Belt}, this method smoothly transitions from an upper limit to a central interval for the signal $\mu$ as one moves from a null result (compatible with background-only expectation, left panel) to a non-null result (right panel).


\bibliographystyle{JHEP}
\bibliography{references}
\end{document}